%% file: template.tex
\PassOptionsToPackage{unicode}{hyperref}
\PassOptionsToPackage{hyphens}{url}
\PassOptionsToPackage{dvipsnames,svgnames,x11names}{xcolor}
\documentclass[
  11pt]{article}

\usepackage{amsmath,amssymb}
\usepackage[table]{xcolor}
\usepackage{dcolumn}
\usepackage{tabularx}
\newcolumntype{Y}{>{\centering\arraybackslash}X}
\newcommand{\sym}[1]{\ifmmode^{#1}\else\(^{#1}\)\fi}
\usepackage{listings}
\usepackage{xcolor}

\usepackage{iftex}
\ifPDFTeX
  \usepackage[T1]{fontenc}
  \usepackage[utf8]{inputenc}
  \usepackage{textcomp} 
\else 
  \usepackage{unicode-math}
  \defaultfontfeatures{Scale=MatchLowercase}
  \defaultfontfeatures[\rmfamily]{Ligatures=TeX,Scale=1}
\fi
\usepackage{lmodern}
\ifPDFTeX\else  
\fi
\IfFileExists{upquote.sty}{\usepackage{upquote}}{}
\IfFileExists{microtype.sty}{
  \usepackage[]{microtype}
  \UseMicrotypeSet[protrusion]{basicmath} 
}{}
\usepackage{xcolor}
\makeatletter
\ifx\paragraph\undefined\else
  \let\oldparagraph\paragraph
  \renewcommand{\paragraph}{
    \@ifstar
      \xxxParagraphStar
      \xxxParagraphNoStar
  }
  \newcommand{\xxxParagraphStar}[1]{\oldparagraph*{#1}\mbox{}}
  \newcommand{\xxxParagraphNoStar}[1]{\oldparagraph{#1}\mbox{}}
\fi
\ifx\subparagraph\undefined\else
  \let\oldsubparagraph\subparagraph
  \renewcommand{\subparagraph}{
    \@ifstar
      \xxxSubParagraphStar
      \xxxSubParagraphNoStar
  }
  \newcommand{\xxxSubParagraphStar}[1]{\oldsubparagraph*{#1}\mbox{}}
  \newcommand{\xxxSubParagraphNoStar}[1]{\oldsubparagraph{#1}\mbox{}}
\fi
\makeatother

\usepackage{color}
\usepackage{fancyvrb}

\DefineVerbatimEnvironment{Highlighting}{Verbatim}{commandchars=\\\{\}}
\usepackage{framed}
\definecolor{shadecolor}{RGB}{241,243,245}

\usepackage{longtable,booktabs,array}
\usepackage{calc} 
\usepackage{etoolbox}
\makeatletter
\patchcmd\longtable{\par}{\if@noskipsec\mbox{}\fi\par}{}{}
\makeatother
\IfFileExists{footnotehyper.sty}{\usepackage{footnotehyper}}{\usepackage{footnote}}
\makesavenoteenv{longtable}
\usepackage{graphicx}
\makeatletter
\newsavebox\pandoc@box
\newcommand*\pandocbounded[1]{
  \sbox\pandoc@box{#1}%
  \Gscale@div\@tempa{\textheight}{\dimexpr\ht\pandoc@box+\dp\pandoc@box\relax}%
  \Gscale@div\@tempb{\linewidth}{\wd\pandoc@box}%
  \ifdim\@tempb\p@<\@tempa\p@\let\@tempa\@tempb\fi
  \ifdim\@tempa\p@<\p@\scalebox{\@tempa}{\usebox\pandoc@box}%
  \else\usebox{\pandoc@box}%
  \fi%
}
\def\fps@figure{htbp}
\makeatother

\usepackage{graphicx}
\usepackage{bbm}
\usepackage{amssymb}
\usepackage{amsmath}
\usepackage[nomarginpar]{geometry}
\usepackage{setspace}
\usepackage{natbib}
\usepackage{lscape}
\usepackage{rotating}
\usepackage{tabularx, calc}
\usepackage{threeparttable}
\usepackage{picinpar}  
\usepackage{longtable}
\usepackage{ae}
\usepackage{float}
\usepackage{morefloats}
\usepackage{palatino}
\usepackage{hyperref}
\usepackage{color}
\usepackage{adjustbox}
\usepackage{booktabs} 
\usepackage{ulem}
\usepackage{ragged2e}
\usepackage{changepage}
\usepackage{longtable}
\usepackage{setspace}

\makeatletter
\@ifpackageloaded{caption}{}{\usepackage{caption}}
\AtBeginDocument{%
\ifdefined\contentsname
  \renewcommand*\contentsname{Table of contents}
\else
  \newcommand\contentsname{Table of contents}
\fi
\ifdefined\listfigurename
  \renewcommand*\listfigurename{List of Figures}
\else
  \newcommand\listfigurename{List of Figures}
\fi
\ifdefined\listtablename
  \renewcommand*\listtablename{List of Tables}
\else
  \newcommand\listtablename{List of Tables}
\fi
\ifdefined\figurename
  \renewcommand*\figurename{Figure}
\else
  \newcommand\figurename{Figure}
\fi
\ifdefined\tablename
  \renewcommand*\tablename{Table}
\else
  \newcommand\tablename{Table}
\fi
}
\@ifpackageloaded{float}{}{\usepackage{float}}
\floatstyle{ruled}
\@ifundefined{c@chapter}{\newfloat{codelisting}{h}{lop}}{\newfloat{codelisting}{h}{lop}[chapter]}
\floatname{codelisting}{Listing}

\makeatother
\makeatletter
\@ifpackageloaded{caption}{}{\usepackage{caption}}
\@ifpackageloaded{subcaption}{}{\usepackage{subcaption}}
\makeatother

\usepackage[]{natbib}
\usepackage{bookmark}

\IfFileExists{xurl.sty}{\usepackage{xurl}}{} 
\hypersetup{
  pdftitle={Paper template},
  pdfauthor={Hyoungchul Kim; Author 2},
  pdfkeywords={3 to 6 keywords},
  colorlinks=true,
  linkcolor={cyan},
  filecolor={Maroon},
  citecolor={cyan},
  urlcolor={cyan},
  pdfcreator={LaTeX via pandoc}}

\title{Paper template}
\author{Sean Cao \and Jiang Wei \and Hui Xu}
\date{\today}

\begin{document}
\def\spacingset#1{\renewcommand{\baselinestretch}%
{#1}\small\normalsize} \spacingset{1}


\date{
\vspace{1em}  
~
}
\title{Seeing the Goal, Missing the Truth: Human Accountability for AI Bias \thanks{We thank seminar participants at the University of Connecticut, the University of Maryland, Lancaster University, the University of California -Riverside, Hitotsubashi University, Waseda University, Singapore Management University, Nanyang Technological University, Tulane University, Cheung Kong Graduate School of Business (CKGSB), Peking University, the University of Central Florida, Baruch College, Stony Brook University, UBS, and the China International Conference in Finance (CICF) for helpful feedback. Sean Cao gratefully acknowledges support from the Smith AI Initiative for Capital Markets Research at the University of Maryland.}}
\author{
Sean Cao\thanks{Robert H. Smith School of Business, University of Maryland, College Park. Email: \href{mailto:scao824@umd.edu}{scao824@umd.edu}}\\
University of Maryland\\
\and Wei Jiang\thanks{Goizueta Business School, Emory University. Email: \href{mailto:wei.jiang@emory.edu}{wei.jiang@emory.edu}}\\
Emory University\\
\and Hui Xu\thanks{Lancaster University Management School, Lancaster University. Email: \href{mailto:h.xu10@lancaster.ac.uk}{h.xu10@lancaster.ac.uk}}\\
Lancaster University
}
\maketitle

\bigskip
\bigskip
\begin{abstract}
This research explores how human-defined goals influence the behavior of Large Language Models (LLMs) through purpose-conditioned cognition. Using financial prediction tasks, we show that revealing the downstream use (e.g., predicting stock returns or earnings) of LLM outputs leads the LLM to generate biased sentiment and competition measures, even though these measures are intended to be downstream task–independent. Goal-aware prompting shifts these intermediate measures toward the disclosed downstream objective, producing in-sample overfitting. Specifically, purpose leakage improves performance on data prior to the LLM’s knowledge cutoff, but provides no advantage after the cutoff. This bias is strong enough that regularization of prompt instructions cannot fully address this form of overfitting. We further show that the bias can arise from users' unintentional conversational context that hints at the purpose. Overall, we document that AI bias due to ``seeing the goal'' is not an algorithmic flaw, but stems from human accountability in research design. 

\end{abstract}

\bigskip
\noindent%
\vfill

\newpage

\begin{center}
    {\Large Seeing the Goal, Missing the Truth: Human Accountability for
AI Bias}
\\ \ \\ \ \\

\textbf{ABSTRACT}
\end{center}

\begin{quote}
This research explores how human-defined goals influence the behavior of Large Language Models (LLMs) through purpose-conditioned cognition. Using financial prediction tasks, we show that revealing the downstream use (e.g., predicting stock returns or earnings) of LLM outputs leads the LLM to generate biased sentiment and competition measures, even though these measures are intended to be downstream task–independent. Goal-aware prompting shifts these intermediate measures toward the disclosed downstream objective, producing in-sample overfitting. Specifically, purpose leakage improves performance on data prior to the LLM’s knowledge cutoff, but provides no advantage after the cutoff. This bias is strong enough that regularization of prompt instructions cannot fully address this form of overfitting. We further show that the bias can arise from users' unintentional conversational context that hints at the purpose. Overall, we document that AI bias due to ``seeing the goal'' is not an algorithmic flaw, but stems from human accountability in research design.  
\end{quote}
\doublespacing

\section{Introduction}\label{sec-intro}

In organizational settings, the disclosure of downstream use can alter the nature of an intermediate task. Consider a human assistant asked to summarize interview transcripts post hoc. If told that the summaries will be used to evaluate recruiting effectiveness, the assistant may emphasize the strengths and ``bright spots'' of candidates who ultimately receive offers, while downplaying uncertainty, mixed signals, or unfavorable assessments. The resulting summaries support a narrative of successful recruitment, yet are less faithful to the underlying interviews. This behavioral shift reflects rational adaptation to perceived downstream purpose rather than error or lack of effort. More broadly, when intermediate tasks are conditioned on their eventual application, neutrality can be compromised even in the absence of explicit incentives.

Purpose-conditioned cognition is a well-documented human trait.\footnote{Related ideas appear in the literature on motivated reasoning and belief distortion \citep{BenabouTirole2016}, framing effects in decision-making \citep{TverskyKahneman1981} and the limits of attention and interpretation in complex environments \citep{mullainathan2008coarse}. While the terminology differs across fields, these studies share the insight that cognition and interpretation are systematically shaped by perceived objectives and contextual framing.} Individuals often adjust interpretation and judgment in response to perceived goals or downstream use, even without explicit incentives; and such behavior is difficult to eliminate. As a result, human-generated intermediate outputs are understood to embed goal-related distortions. On the other hand, one might expect algorithmic systems, including large language models (LLMs), to be immune from human biases. This presumption motivates a natural question: does purpose-conditioned behavior also arise in LLMs? In other words, if we restrict AI to ``assistant'' roles, does this separation from the final decision task ensure neutral and unbiased intermediate outputs?

Our research design addresses this question by holding the input text, model, and scoring task fixed, while varying only whether the downstream use of the output is disclosed. Specifically, we prompt two large language models (ChatGPT and Gemini) to generate numerical measures of sentiment and competition intensity from the same source information (earnings call transcripts) but under two prompt conditions. In the goal-blind condition, the model is asked to produce a score without any reference to its eventual use. In the goal-aware condition, the model is informed that its output will be used to predict future stock returns or earnings. This design therefore isolates the effect of disclosing downstream use on the model’s intermediate outputs.

Disclosing downstream use materially alters both the statistical properties and the economic content of LLM-generated sentiment scores. In standard portfolio-sorting tests, goal-aware sentiment produces substantially larger return spreads than goal-blind sentiment prior to the model's knowledge cutoff. This pattern suggests that the model conditions its intermediate outputs on the disclosed objective and implicitly re-weights return-relevant patterns learned during training. The relative advantage of the goal-aware regime disappears after the knowledge cutoff, indicating that the earlier performance gap is tied to information availability rather than a stable improvement in signal quality.  

This comparison of the two prompt regimes is further confirmed in \cite{fama1973risk} return-prediction regressions. Intermediate variables generated under goal-aware prompts exhibit stronger incremental predictive power than those from the goal-blind regime in the pre-cutoff period, both in coefficient significance and in out-of-sample $R^2$. Once the model no longer has access to future information beyond the cutoff, this advantage dissipates. Together, these results are consistent with objective-conditioned output adjustment rather than a structurally more informative sentiment measure.  In fact, the enhanced goodness-of-fit prior to knowledge cutoff even slightly degrades out-of-sample model generalization. On the surface, the pre-cutoff inflation and post-cutoff attenuation of goal-aware advantage look like the p-hacking and data-mining problems documented in the anomalies literature.\footnote{See, e.g., \cite{HarveyLiuZhu2016,HouXueZhang2020,McLeanPontiff2016}} But the mechanism is different. P-hacking distorts because researchers cycle through specifications and report the ones that survive conventional thresholds.  In our setting, the inflation occurs within a single forward pass. The model shifts its outputs toward return-predictive patterns as soon as it is told what those outputs are for. Pre-registration and multiple-testing corrections cannot address this, because there is only one test.

One might expect that a carefully worded instruction can undo what disclosure triggers, but we further show that explicit disclosure is not required. ``Subtle'' clues (e.g., asking the LLM whether sentiment scores predict stock returns, with no instruction linking that context to the scoring task) generate largely the same outcome.  Even an explicit directive asking the model to minimize any reliance on outcomes beyond the measurement window, effectively auditing its own reasoning for hindsight, merely attenuates the bias instead of eliminating it. Once information is encoded in model weights, prompt-level instructions cannot selectively suppress it at inference time.

The seeming goal-conditioning behavior of a cold-blooded algorithm has its roots in how AI systems are trained and deployed. Large language models are optimized to generate outputs that satisfy the objective implied by the prompt, conditional on their learned representations. Disclosing downstream use alters this implicit objective, encouraging responses that align with anticipated evaluation criteria rather than with neutral processing of the input. Put differently, while a goal-blind prompt positions the LLM as a neutral sensor, a goal-aware prompt transforms it into a strategic tuner. This behavior does not require intent or explicit embedding of forward-looking information. Instead, it reflects the model’s reliance on correlations learned during training and its sensitivity to contextual cues about what constitutes a desirable output. As a result, goal awareness can improve in-sample alignment while reducing robustness when the information environment changes.

Our study contributes to the AI and LLM methodology literature (reviewed in Appendix \ref{sec-lit}) in two distinct ways. First, existing research largely attributes AI bias to algorithmic limitations. Look-ahead bias and memorization, for example, are commonly traced to models’ access to unintended training data \citep*{glasserman2023assessing,sarkar2024lookahead,lopez2025memorization,he2025chronologically,cao2025llms}. Moving beyond these training data-centric issues, AI has also been criticized for potential biases in exploration strategies and model weights, which may be related to ethical concerns \citep{fedyk2024ai} and preference alignment \citep{ouyang2024ai,hirshleifer2025social,NBERw34745}. Our analysis shifts the focus from these machine-level limitations to human use of machine. We show that human disclosure of downstream task (e.g., return and earnings predictions) reshapes intermediate outputs (e.g., sentiment and competition measures) in ways that inflate the downstream task performance. Ultimately, this leads to impressive in-sample results that fail to generalize out-of-sample.

The second contribution of the paper is to isolate goal awareness from other AI biases, which naturally emerge when LLMs are deployed directly on end tasks such as return forecasting. A commonly proposed safeguard is to limit LLMs to intermediate functions, analogous to a research assistant, while reserving final predictions and decisions for human judgment. Our results show that even at the intermediate-measurement stage, disclosure of the downstream objective can introduce additional, goal-conditioned distortion in the generated signals. The conversation history shows that a model with encountered relevant context infers the downstream objective and adjusts without any direct instruction. Prompt-level regularization can reduce but not fully remove this bias, since the underlying knowledge is distributed across model parameters rather than stored in a form that instructions can selectively override.

More broadly, the analysis clarifies the distinction between using AI as a task agent that optimizes directly toward an announced objective and using it as a measurement tool that produces inputs for subsequent human evaluation. Prompt and workflow designs that deliberately reduce objective conditioning can therefore mitigate machine-driven bias. The mechanism we uncover thus bears an analogy to the phenomenon of ``AI sycophancy'' discussed in recent work, e.g., \cite{SharmaEtAl2023Sycophancy}. In both cases, large language models adjust their outputs in response to contextual cues about what constitutes a desirable response, rather than adhering to a task-invariant notion of correctness or neutrality.\footnote{In sycophancy, this adjustment manifests as alignment with a user’s stated beliefs or preferences; in our setting, it appears as alignment with an inferred evaluation objective.} The underlying commonality is not intent, but sensitivity to signals about reward or approval embedded in the prompt. As a result, both phenomena illustrate how models trained to be helpful and aligned can deviate from neutral information processing when the prompt conveys implicit incentives, even in the absence of explicit instructions or forward-looking information.\footnote{In computer science, related concerns are discussed under reward hacking, specification gaming, and objective misgeneralization, where models optimize inferred proxy objectives rather than the intended task. Our setting requires no change in model parameters or rewards; the distortion arises solely from human framing at inference time.} A design choice that weakens alignment with a stated downstream goal can improve statistical validity and out-of-sample reliability.  

In a nutshell, intermediate variables constructed with AI assistance should, whenever feasible, be generated under prompts agnostic to downstream use and evaluated with strict out-of-sample tests. Treating LLMs as neutral measurement devices requires not only holding inputs and models fixed, but also constraining the informational environment to limit objective inference by the model beyond the intrinsic requirements of the task. In this sense, the challenge is not primarily one of algorithmic bias, but of human accountability in how objectives, context, and evaluation criteria are built in the system.

\section{Experimental Design and Data}
\subsection{LLM Prompt and Scoring}
Our objective is to examine whether large language models (LLMs) systematically adjust their outputs when they are informed about the downstream task for which those outputs will ultimately be used. We focus on two economically important forecasting applications: monthly stock returns and quarterly earnings per share (EPS). In both settings, the firm’s most recent earnings call transcript serves as the sole input to the LLM. Rather than asking the model to directly predict economic outcomes, we instruct it to generate an intermediate score that is subsequently used in a simple predictive regression.
\begin{itemize}
    \item Return prediction task. For each calendar month $t$, the model is prompted with the most recent earnings call transcript available at $t-1$ to generate a continuous sentiment score, ranging from -1 to 1, summarizing the firm’s business sentiment for month $t$. We then use this month-$t$ sentiment score to predict the firm’s stock returns in month $t$.
    
    \item Earnings prediction task. For each fiscal quarter $t$, the model is prompted with the most recent earnings call transcript available at $t-1$ to generate a competition score, ranging from -1 to 1, which captures the intensity of competitive pressure the firm faces in quarter $t$. We then use this quarter-$t$ competition score to predict the firm’s earnings realized in quarter $t$.

\end{itemize}

A central feature of the experimental design is that the LLM does not directly forecast stock returns or earnings. Instead, it produces an \textit{intermediate} construct (i.e., sentiment or competitive intensity), which we subsequently map to economic outcomes using standard econometric models. This design choice allows us to test whether the LLM adjusts these intermediate scores as if it were strategically responding to the knowledge of their downstream use. For each task, we construct two prompts that are otherwise identical in wording and structure, differing only in whether the prompt explicitly discloses the ultimate use of the generated score.

\begin{enumerate}
    \item Goal-Blind Prompt (control): The LLM is asked to generate the score without being told that it will be used in a forecasting regression.
    \item Goal-Aware Prompt (treatment): The LLM is informed that the score it produces will later be used as an explanatory variable in a regression to predict either stock returns or earnings.
\end{enumerate}

This minimalist treatment variation allows us to cleanly isolate whether, and through what channels, an LLM alters its outputs once it becomes aware of the user’s ultimate objective. Conceptual concerns arise from anticipatory behavior, in which the model conditions its responses on inferred downstream use rather than on the stated prediction task alone; reward hacking, whereby the model optimizes for the perceived evaluation criterion instead of the intended informational objective; and goal misalignment, in which the model’s internal optimization departs from the user’s declared constraints. Importantly, these issues are examined within an economically meaningful forecasting environment, allowing deviations in model behavior to be traced to distortions in information processing rather than an abstract alignment failure.

To illustrate, the following text shows the goal-blind prompt used to generate sentiment scores:
\begin{quote}
\small
\texttt{"For the following tasks, all dates are expressed in the format MM/DD/YYYY (month/day/year). \\
Below is the earnings call transcript of }\textit{\{ticker\}}\texttt{. Please provide a continuous sentiment score in [-1, 1] about the firm's business sentiment for the month ending on }\textit{\{date\}}\texttt{. \\
Provide a precise numerical answer. Format as a JSON object with the following fields: \\
- answer: The precise numerical answer to the question. No strings.
\\}\textit{\{the firm's earnings call transcript\}}\texttt{."}
\end{quote}

The corresponding goal-aware version differs only by the addition of a single sentence that discloses the downstream use of the output, namely the predictive task.

\begin{quote}
\small
\texttt{"For the following tasks, all dates are expressed in the format MM/DD/YYYY (month/day/year). \\
Below is the earnings call transcript of }\textit{\{ticker\}}\texttt{. Please provide a continuous sentiment score in [-1, 1] about the firm's business sentiment for the month ending on }\textit{\{date\}. }\textbf{The sentiment score later will be used as an explanatory variable in a regression to predict the monthly stock  returns ending on {\{\textit{date}\}}}\texttt{. \\
Provide a precise numerical answer. Format as a JSON object with the following fields: \\
- answer: The precise numerical answer to the question. No strings.
\\}\textit{\{the firm's earnings call transcript\}}\texttt{."}
\end{quote}

Aside from this one sentence, all instructions, including input data, output format, and numerical constraints, remain identical across the two prompts and processes.\footnote{The full prompts are also detailed in Appendix \ref{sec-prompts}.}

\subsection{Data and LLM models}

Our sample consists of S\&P 500 firms over the period from January 2022 to December 2024. Earnings call transcripts are obtained from Capital IQ, which provides standardized, time-stamped transcripts for publicly listed firms. Monthly stock return data are drawn from CRSP, and accounting information, including earnings per share (EPS), is sourced from Compustat. We further use CRSP and Compustat data to construct standard firm-level control variables employed in our predictive regressions, such as the book-to-market ratio and firm size.

To generate sentiment and competition scores, our baseline experiment uses the GPT-4o-mini model. GPT-4o-mini (``o” for ``omni”) is a lightweight, cost-efficient language model designed for focused inference tasks with low latency. According to OpenAI, GPT-4o-mini is produced via distillation from a larger frontier model (GPT-4o), allowing it to replicate much of the larger model’s behavior at substantially lower computational cost.  Critically for our experimental design, GPT-4o-mini has a fixed knowledge cutoff of October 1, 2023. As a result, the model does not have access to information that occurs after this date. This feature is central to our analysis, as it allows us to cleanly separate changes in predictive performance driven by prompt design and goal awareness from those arising from direct access to future information.

As a robustness test, we repeat the same analyses using Gemini 2.5, a large language model developed by Google with a well-documented knowledge cutoff date of January 2025.\footnote{Google does not disclose an exact knowledge cutoff date for Gemini 2.5, indicating only January 2025. We therefore set the cutoff to January 1, 2025. Given that our downstream prediction task is based on monthly returns measured at the last trading day of each month, this convention is unlikely to materially affect our results.}  

\subsection{Evaluation Metrics}\label{sec-eval_metrics}
We first evaluate the economic relevance of GPT-generated sentiment scores using portfolio-sorting tests that are standard in the asset-pricing literature. In each period, firms are sorted into quintiles based on their GPT-produced sentiment scores, and we form an equally weighted zero-investment long–short portfolio that buys firms in the highest quintile and sells firms in the lowest quintile. Portfolio performance is measured using average excess returns. The analysis is conducted separately for goal-aware and goal-blind scores. Comparing portfolio performance across the two prompt designs allows us to assess whether revealing the ultimate prediction task leads to economically stronger trading signals, and whether the relative performance of goal-aware versus goal-blind scores changes after the cutoff date.


We evaluate predictive performance using two complementary approaches: \cite{fama1973risk} predictive regressions and genuine out-of-sample forecasting. The first approach tests whether LLM-generated scores significantly predict returns or earnings in the full sample and how this relationship changes around the model's knowledge cutoff. The second assesses practical forecasting accuracy by simulating real-time predictions on unseen data. Together, these methods provide a robust validation: the regressions establish statistical significance of the predictive variable, while the out-of-sample exercise determines whether this significance translates into reliable practical performance.

We use standard cross-sectional predictive regressions to examine statistical predictability. For stock returns, we implement the \cite{fama1973risk} methodology and estimate monthly cross-sectional regressions of the form:
\begin{align}
    R_{i,t} = & \beta_{1,t} \mathrm{Score}_{i,t-1} \times \mathrm{Pre\text{-}Cutoff}_{t}+ \beta_{2,t} \mathrm{Score}_{i,t-1} \times \mathrm{Post\text{-}Cutoff}_{t}+ \nonumber \\
    & \beta_{3,t} \mathrm{Diff}_{i,t-1} \times \mathrm{Pre\text{-}Cutoff}_{t}+ \beta_{4,t} \mathrm{Diff}_{i,t-1} \times \mathrm{Post\text{-}Cutoff}_{t} + \alpha_{t} + \epsilon_{i,t}, \label{eq:fmb}
\end{align}
where $R_{i,t}$ denotes the excess return of firm $i$ in month $t$. \textit{Score}$_{i,t-1}$ is the sentiment score generated by the goal-blind GPT prompt using the most recent earnings call transcript available at $t-1$. \textit{Diff}$_{i,t-1}$ is the difference between the sentiment scores produced by the goal-aware and goal-blind prompts. Because the sentiment score does not have a natural scale, raw values may not be directly comparable across firms, industries, or time periods. For this reason we transform both sentiment scores into percentiles across all firms within each year–month so that \textit{Diff}$_{i,t-1}$ variable captures the gap in the percentile values of scores generated under goal-aware and goal-blind scores.  The time-period fixed effect $\alpha_t$ is included to absorb aggregate time series shifts in returns.

We interact both \textit{Score}$_{i,t-1}$ and \textit{Diff}$_{i,t-1}$ with indicator variables \textit{Pre-Cutoff}$_{t}$ and \textit{Post-Cutoff}$_{t}$, which equal one if month $t$ occurs before or after the LLM’s knowledge cutoff date, respectively. This specification allows predictive coefficients to differ across the two periods and is designed to isolate the role of goal awareness. When the LLM is goal aware, it may leverage patterns, associations, and correlations embedded in the full training sample available that are correlated with future outcomes. Because such information is internalized during training and cannot be selectively ``unlearned'' at inference time, goal awareness can thus induce outputs that are implicitly forward-looking relative to the evaluation period, leading to stronger predictive performance prior to the cutoff. Such behavior does not require the LLM to explicitly access forward-looking information in making predictions.  Once the evaluation period extends beyond the model’s knowledge cutoff, however, these internalized correlations become equally stale, and the performance advantage of goal-aware scores over goal-blind scores should dissipate or even reverse.

Accordingly, any differential predictive power attributable to goal awareness should be concentrated in the pre-cutoff period and attenuated in the post-cutoff period. For this reason, we expect $\beta_{1,t}$ and $\beta_{2,t}$ in Equation (\ref{eq:fmb}) to be positive if the intermediate variable is informative. Moreover, the two coefficients should be of similar magnitude, as knowledge of future returns prior to the cutoff date should not matter under goal blindness. By the same reasoning, $\beta_{4,t}=0$ since the lack of future-relevant information renders any advantage of goal awareness nil after the cutoff. Finally, $\beta_{3,t}$ is the key coefficient of interest. Under the null, $\beta_{3,t}=0$ if goal awareness does not change the way the LLM generates outputs; under the alternative, $\beta_{3,t}$ may be positive if disclosure of the downstream use of the output motivates the LLM to produce intermediate measures that better align with the ultimate task.

For earnings outcomes, we adopt an analogous specification, replacing monthly excess returns with quarterly earnings per share (EPS). We use diluted EPS excluding extraordinary items from Compustat, following common practice in the literature. In this setting, \textit{Score}$_{i,t-1}$ corresponds to the competition score generated by the goal-blind prompt, transformed into percentiles within the same 4-digit SIC industry $\times$ year cohort. The difference measure captures the gap between the scores generated under the goal-blind and goal-aware prompts. This structure, which is parallel to our earlier exercise on stock returns, facilitates multiple validation by allowing us to test how goal awareness affects predictability across both financial and accounting settings.

A complementary approach to the cross-sectional predictive regression is assessing genuine, out-of-sample forecasting performance. At the beginning of each forecast period, $T$, we estimate the regression model on all data available through $T-1$ and generate predictions for outcomes at $T$; this estimation window is then expanded recursively to include additional observations before each new forecast, ensuring that every prediction is made using the maximum available history.\footnote{This procedure is equivalent to supervised learning with an expanding training window: at each forecast period $T$, the model is trained on all observations $\{1,\ldots…,T-1\}$ and evaluated on the held-out observation at $T$. The training set grows monotonically with each step, and no future data ever enters the training set prior to the corresponding forecast, a design that strictly preserves temporal holdout and precludes look-ahead bias.} The unit of observation is monthly for stock returns and quarterly for earnings. For stock returns prediction, this amounts to the following predictive regression:
\begin{align}
    R_{i,t} = \alpha +  \gamma \, \mathrm{Score}_{i,t-1} +\epsilon_{i,t}, \forall  t\le T-1.
\end{align}
We then use the estimated coefficients $\{\hat{\alpha}, \hat{\gamma}\}$ to predict stock returns $R_{i,T}$ based on sentiment scores observed at $T-1$. We estimate this regression separately using scores generated from the goal-aware and goal-blind prompts. Earnings predictions follow an analogous specification, with one modification: rather than using the level of EPS directly, we replace it with the year-over-year growth rate of same-quarter EPS, 
defined as $\frac{\text{\textit{EPS}}_{i,T}}{\text{\textit{EPS}}_{i,T-4}}$. This adjustment accounts for the well-documented seasonality and serial correlation in earnings.

Once we obtain model-based forecasts, we construct a forecast accuracy measure analogous to the out-of-sample ($OOS$) $R^{2}$ \citep{campbell2008predicting}:
\begin{equation}
R^{2}_{OOS, i, T}=1-\frac{(y_{i,T}-\hat{y}_{i,T})^{2}}{(\bar{y}_{T-1}-y_{i,T})^{2}},
\end{equation}
where $y_{i,T}$ is the realized outcome, $\hat{y}_{i,T}$ is the model-based forecast, and $\bar{y}_{T-1}$ is  the benchmark forecast—specifically, the simple historical cross-sectional mean of the outcome variable across all firms, calculated using data available through period $T-1$.

With the forecast accuracy measure at the firm-period level, we are able to assess their performance in relation to goal-awareness and interacted with the LLM knowledge cutoff date.  More specifically, we estimate the following panel regression:
\begin{align} \label{eq:r2reg}
    R^{2}_{OOS, i, T} = \theta_1 \mathrm{Goal\text{-}Aware}_{i,T}\times\mathrm{Post\text{-}Cutoff}_{T}+\theta_2 \mathrm{Goal\text{-}Aware}_{i,T}+\theta_3 \mathrm{Post\text{-}Cutoff}_{T}+\mu_{i}+\nu_{T}+\epsilon_{i,T}
\end{align}
where \textit{Goal-Aware}$_{i,T}$ is an indicator equal to one (zero) if $R^{2}_{OOS, i, T}$ results from the goal-aware (goal-blind) regime. $\mu_{i}$ and $\nu_{T}$ denote firm and time fixed effects, respectively. The coefficient $\theta_1$ is of central interest, as it captures the change in incremental forecasting performance of goal-aware relative to goal-blind scores before and after the model’s knowledge cutoff date. A negative $\theta_1$ indicates that the incremental forecasting performance of goal-aware prompt weakens after the knowledge cutoff.

\section{Findings} \label{sec:findings}
\subsection{Sentiment Scores and Stock Returns Prediction}
\subsubsection{Portfolio Sorts and Return Performance}
\begin{figure}[htbp!]
\centering
\includegraphics[width=1.0\textwidth]{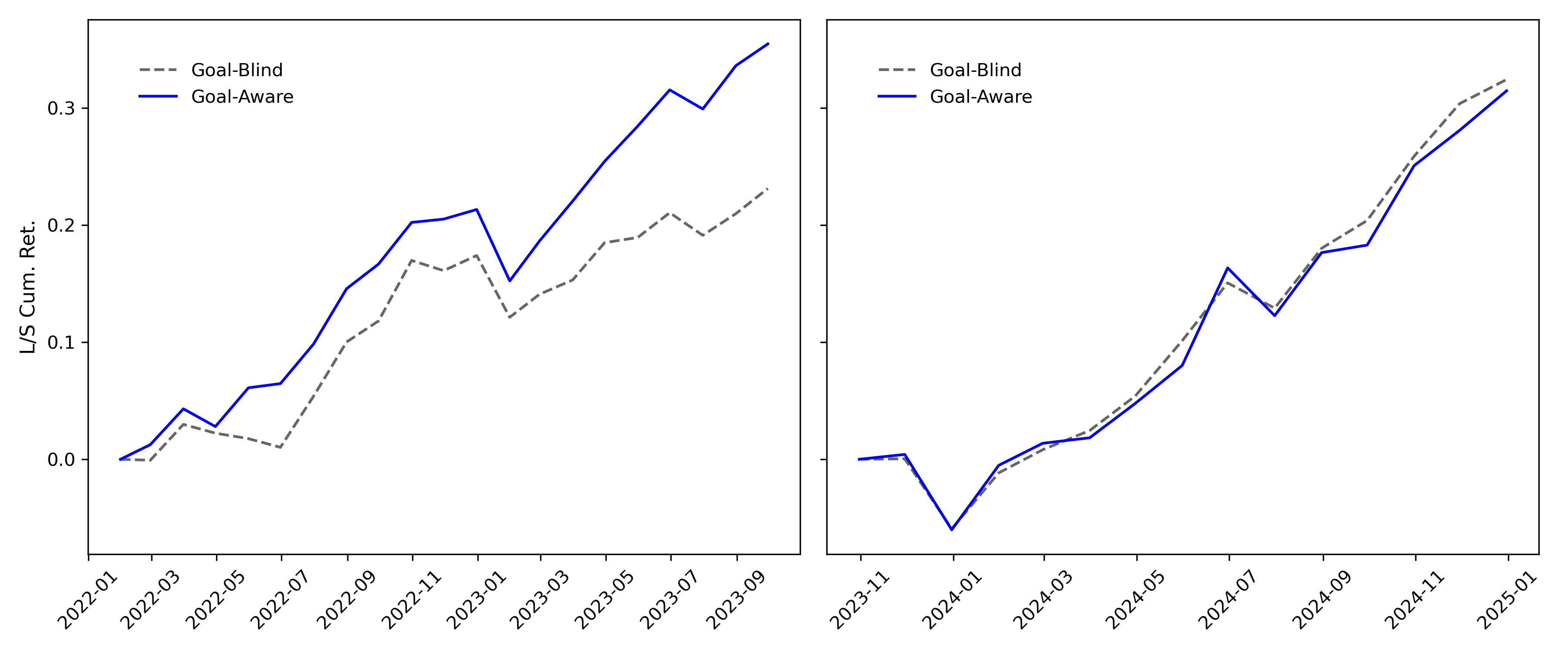}
\caption{Cumulative Long–Short Portfolio Returns from Goal-Aware and Goal-Blind Sentiment Scores}
\caption*{This figure plots the cumulative returns to long–short portfolios constructed using GPT-generated sentiment scores. Each month, firms are sorted into quintiles based on their sentiment scores, separately for the goal-blind and goal-aware prompts. The portfolio goes long the highest-sentiment quintile and short the lowest-sentiment quintile, and cumulative returns are computed over time. To facilitate comparison of pre- and post-cutoff performance, cumulative returns are normalized to start at zero at the cutoff. The figure illustrates how the long–short strategy based on goal-aware sentiment evolves relative to the strategy based on goal-blind sentiment before and after the knowledge cutoff.}
\label{fig:ret_sentiment_ls}
\end{figure}

We use GPT as the primary LLM model for illustration, followed by replicated findings on Gemini in Section \ref{sec:gemini}.  We begin by examining the economic implications of GPT-generated sentiment scores using portfolio-sorting tests. \textcolor{black}{The purpose of this exercise is to verify that the sentiment scores have predictive content for stock returns. Although return forecasting is not a central objective of this study, the presence of a predictive variable is required to evaluate how the LLM’s output changes when the model is made aware of the downstream predictive task.} Figure \ref{fig:ret_sentiment_ls} plots the cumulative returns to long–short portfolios formed by buying firms in the highest sentiment quintile and selling firms in the lowest sentiment quintile, separately for scores generated under goal-aware and goal-blind prompts. Table \ref{tab:ret_sentiment_ls} reports the corresponding average monthly return spreads for the pre– and post–knowledge cutoff periods.

Prior to the knowledge cutoff, sentiment scores generated under both prompt designs exhibit economically meaningful return predictability. As reported in Table \ref{tab:ret_sentiment_ls}, the long–short portfolio based on goal-aware sentiment earns an average monthly return spread of 1.552\%, while the corresponding spread based on goal-blind sentiment is 1.069\%. Both spreads are statistically significant, indicating that GPT-generated sentiment contains predictive information about future stock returns even when the ultimate objective is not fully communicated to the model. Notably, the goal-aware strategy delivers significantly stronger performance: the difference in High–Low spreads between the two regimes, 0.483 percentage point per month, is statistically significant at the 5\% level. This relative outperformance is evident graphically in Figure \ref{fig:ret_sentiment_ls}, where cumulative returns of the goal-aware portfolio diverge upward from those of the goal-blind portfolio in the period leading up to the cutoff.

After the knowledge cutoff, both strategies continue to generate statistically significant return spreads. The monthly High–Low spread equals 2.269\% for the goal-aware portfolio and 2.239\% for the goal-blind portfolio, with no economically or statistically meaningful difference between the two. Confirming this result, the cumulative return paths in the right panel of Figure \ref{fig:ret_sentiment_ls}, normalizing both portfolios to start from zero at the cutoff date, track each other closely in the post-cutoff period, showing no persistent advantage for the goal-aware strategy. If anything, the goal-blind long–short portfolio performs slightly better. One possible explanation is mild overfitting: making the LLM aware of the downstream objective may induce greater optimization to in-sample patterns, whereas the goal-blind specification generates more stable signals that generalize slightly better in the true out-of-sample (post-cutoff) period.

Taken together, the portfolio evidence shows a contrast around the knowledge cutoff. Before the cutoff, disclosing the downstream prediction task increases the measured economic content of GPT-generated sentiment scores. After the cutoff, this relative difference disappears, even though both prompt designs continue to exhibit statistically significant return predictability in absolute terms. This pattern indicates that the higher pre-cutoff performance of goal-aware sentiment does not arise from superior information extraction. Instead, it is consistent with goal awareness reshaping the distribution of model outputs by incorporating back-tested performance when data from the evaluation period are available during training. The absence of a post-cutoff difference helps identify the source of the pre-cutoff effect. Overall, the results highlight the importance of accounting for goal awareness when interpreting LLM-based measures in empirical studies.

\subsubsection{Predictive Regressions and Out-of-Sample Performance}

The comparative predictive performance of the goal-aware and goal-blind regimes can also be examined within the regression framework described in Section \ref{sec-eval_metrics}. Table \ref{tab:ret_fmb_reg} reports Fama–MacBeth estimates from monthly cross-sectional regressions of excess stock returns on GPT-generated sentiment scores, following Equation \eqref{eq:fmb}. The specification allows the slope coefficients to differ between the goal-aware and goal-blind regimes and across the pre- and post–knowledge cutoff periods.

Across specifications, sentiment scores generated under the goal-blind prompt display economically meaningful and statistically significant predictive power both before and after the knowledge cutoff. The estimated coefficients on the goal-blind score interacted with the pre-cutoff indicator are positive and statistically significant, and their magnitudes remain similar in the post-cutoff period. Formal tests reported in the bottom panel do not reject equality between the pre- and post-cutoff coefficients for the goal-blind score.

By contrast, the incremental predictive content of goal-aware sentiment—captured by the \textit{Diff} measure, exhibits a discontinuity at the knowledge cutoff. In the pre-cutoff period, the coefficient on \textit{Diff} is positive and statistically significant across both specifications, indicating that goal-aware sentiment contains incremental predictive power. The magnitude holds across firm characteristics such as size and book-to-market.  After the knowledge cutoff, the \textit{Diff} coefficient collapses to near-zero. Consistent with this pattern, the bottom panel reports that the equality of the pre- and post-cutoff \textit{Diff} coefficients is rejected at the 5\% significance level. Altogether, the regression evidence closely mirrors the portfolio results. 

\begin{figure}[htbp!]
\centering
\includegraphics[width=1.0\textwidth]{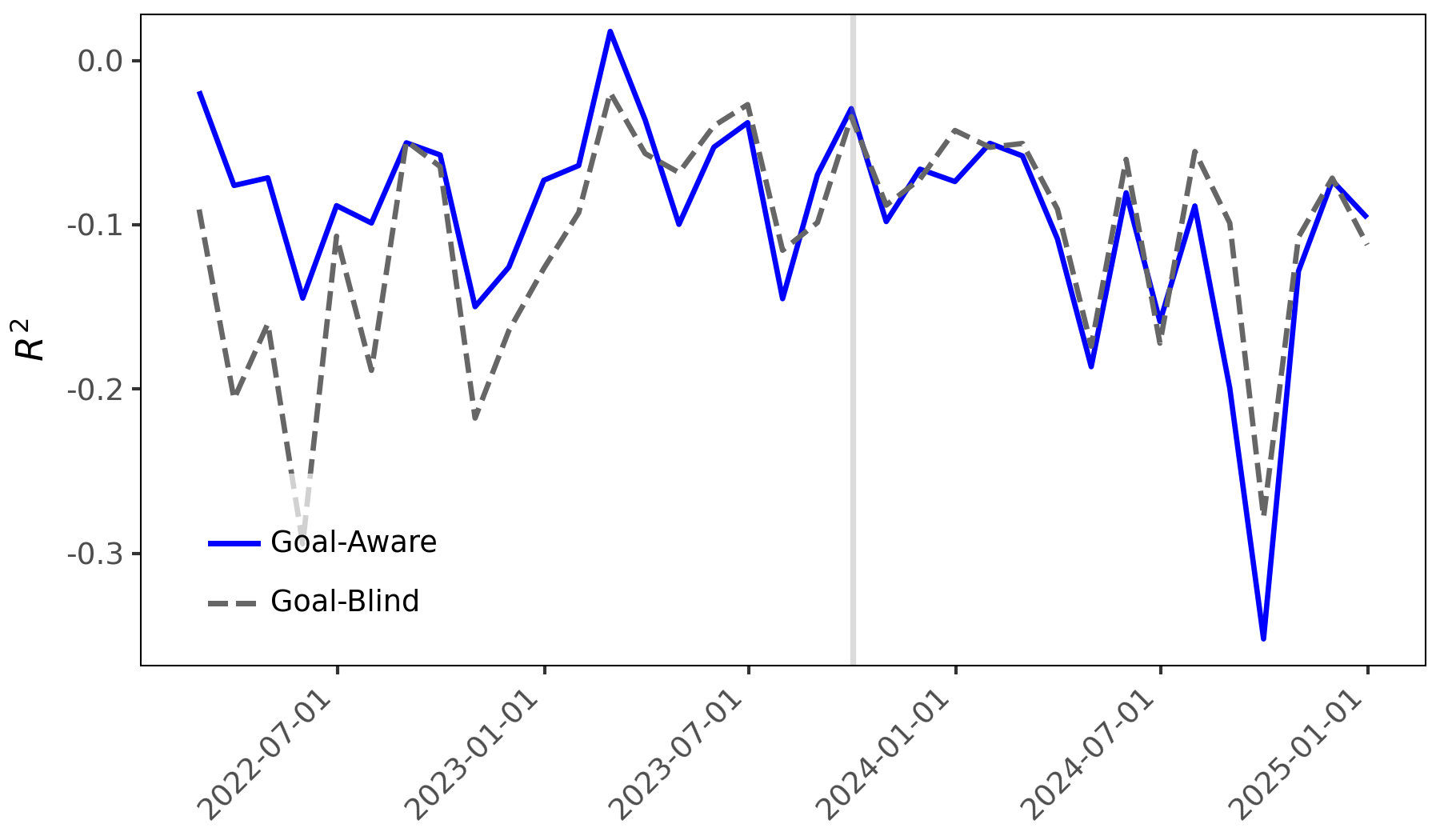}
\caption{Monthly Out-of-Sample Forecast Accuracy Using Goal-Aware and Goal-Blind Sentiment Scores}
\vspace{0.5cm}
\label{fig:ret_oos_r2}
\begin{minipage}{0.95\textwidth}
This figure shows the monthly $R^{2}_{OOS}$ from using goal-aware vs. goal-blind sentiment to forecast stock returns. Each month we run expanding-window regressions, predict next-month returns, compute OOS accuracy relative to the historical mean, and average across firms. The plot highlights how the two prompts track each other before the cutoff and how their predictive behavior evolves once GPT's knowledge becomes stale.
\end{minipage}
\end{figure}

Next we evaluate whether these regression-based differences carry over to out-of-sample forecasting performance as modeled in Equation \eqref{eq:r2reg}. Figure \ref{fig:ret_oos_r2} plots the monthly out-of-sample $R_{OOS}^2$ obtained from forecasting next-month stock returns using sentiment scores generated under goal-aware and goal-blind prompts. In each month $T$, we estimate predictive regressions using an expanding window (where the estimation sample grows to include all available data up to the prior month $T-1$), compute out-of-sample accuracy relative to a simple historical cross-sectional mean benchmark calculated using data available through the prior month $T-1$, and average the resulting out-of-sample goodness-of-fit, $R_{OOS,i,T}^2$, across firms $i$ in each time period $T$.

Figure \ref{fig:ret_oos_r2} reveals a distinct shift in relative predictive accuracy across the knowledge cutoff. Prior to the cutoff, the goal-aware prompt generally outperforms the goal-blind prompt. Following the cutoff, however, this relationship reverses: the performance of the goal-aware prompt deteriorates significantly, falling below that of the goal-blind benchmark. 
Table \ref{tab:ret_oos_r2} formalizes this comparison by regressing monthly $R_{OOS}^2$ values on the interaction between the \textit{Goal-aware} prompt and the \textit{Post-Cutoff} indicators, following the specification in Equation \eqref{eq:r2reg}. Consistent with the visual evidence, the coefficient on the interaction of the indicators is negative and statistically significant across specifications. The magnitude of the estimate implies a meaningful reduction in out-of-sample predictive accuracy for goal-aware sentiment once the model’s knowledge becomes stale. 

The comparative results of out-of-sample predictive performance complement those from portfolio sorting and cross-sectional predictive regressions. While goal-aware sentiment appears to deliver stronger in-sample and pre-cutoff predictive signals, this advantage does not persist when evaluated under a strict out-of-sample framework after the knowledge cutoff. The pattern suggests that the superior pre-cutoff performance of goal-aware sentiment reflects, at least in part, the model’s tendency to condition its outputs on the evaluation objective in ways that do not generalize once the underlying knowledge environment changes. 

Taken together, results from three complementary research designs indicate that making the downstream objective explicit can improve apparent predictive performance in-sample and prior to the knowledge cutoff, but does not improve, or may even degrade out-of-sample generalization. This pattern is consistent with goal awareness inducing objective-conditioned optimization by a LLM that exploits correlations present in the training or evaluation sample at the expense of extracting stable predictive structure. The findings highlight the distinction between economically meaningful signals and prompt-induced optimization artifacts when employing LLM-generated measures in empirical research.

\subsection{Earnings Prediction with Competition Scores}
Similar to our discussion of GPT-generated sentiment scores and their relationship with future stock returns, we first study the ability of competition-threat scores to predict future earnings. Table \ref{tab:eps_fmb} reports Fama–MacBeth estimates from firm-quarter panel regressions of next-quarter earnings per share on GPT-generated competition scores, allowing the predictive slopes to differ between goal-aware and goal-blind regimes and across the pre– and post–knowledge cutoff periods.

As in Table \ref{tab:ret_fmb_reg}, we focus on \textit{Diff}, the difference between goal-aware and goal-blind competition threat scores. Its interactions with \textit{Pre-cutoff} and \textit{Post-cutoff} capture how the goal-aware score's incremental predictive power changes around the knowledge cutoff. Control variables follow \cite{so2013new}: Column (1) includes the previous quarter's truncated EPS (set to zero when negative) and a loss indicator for negative EPS; column (2) includes the full set of controls. The story that emerges is much the same as before: in the pre-cutoff period, the coefficient on \textit{Diff} is negative and statistically significant at the 5\% level across both specifications, consistent with the intuition that fiercer competition depresses earnings. This suggests that the goal-aware LLM tends to assign elevated competition threat scores to firms that subsequently report weaker earnings. Once we move past the knowledge cutoff, this effect vanishes and the \textit{Diff} coefficients drop to near zero. Formal tests reported in the bottom panel reject the equality of the pre- and post-cutoff \textit{Diff} coefficients at conventional significance levels.

\begin{figure}[htbp!]
\centering
\includegraphics[width=1.0\textwidth]{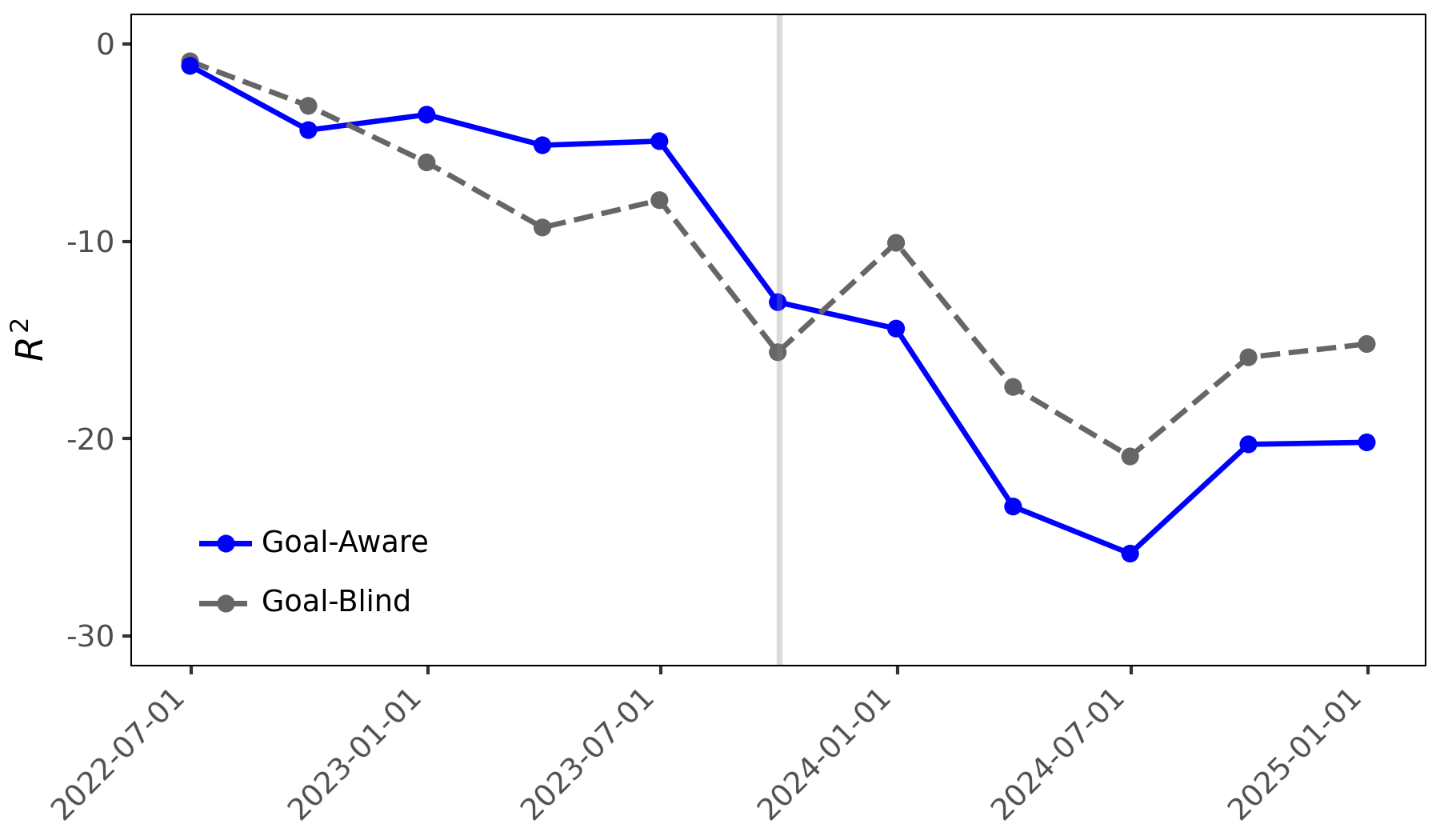}
\caption{Quarterly Out-of-Sample Forecast Accuracy Using GPT-Derived Competition Scores}
\label{fig:eps_r2_oos}
\vspace{0.5cm}
\begin{minipage}{0.95\textwidth}
        We forecast quarterly earnings growth using goal-aware vs. goal-blind competition scores and compute $R^{2}_{OOS}$ each quarter using expanding-window regressions. The plot shows how predictive accuracy evolves around the GPT knowledge cutoff. The incremental predictive power of goal-aware scores deteriorates once the model’s information becomes stale.
\end{minipage}
\end{figure}

We conclude our analysis by evaluating the out-of-sample forecasting power of GPT-generated competition scores for future earnings growth. Figure \ref{fig:eps_r2_oos} plots quarterly average out-of-sample $R^2_{OOS}$ values from expanding-window forecasts of earnings growth (where each forecast uses all available historical data up to the prior quarter) using both goal-aware and goal-blind competition scores. Table \ref{tab:eps_r2_oos} complements this visual evidence with formal regression tests, which confirm that the incremental predictive content of goal-aware scores collapses after the knowledge cutoff.

In sum, the earnings predictions reinforce the central message: While GPT-generated competition scores exhibit economically intuitive relations with future earnings, the incremental predictive content attributable to goal awareness is confined to the pre-cutoff period. This confirms that communicating downstream purpose of the intermediate output may amplify predictive relations in-sample with no out-of-sample generalization.

\section{Mechanism and Subtlety of Goal Awareness} 
\subsection{A theoretical framework}  \label{sec:discussion_reward}

The preceding evidence raises a natural question: why does informing the model of a downstream objective change its output at all? A simple framework provides an answer. Let the LLM's output be an intermediate signal $s$ intended to recover a latent sentiment $\theta$, which predicts returns $y$. Under goal blindness, the model is a \textit{measuring sensor},  minimizes a representation loss and outputs $s = \mathbb{E}[\theta \mid x]$, where $x$ is input text. When told that $s$ will be used to predict returns, the model is an \textit{optimizing solver} and shifts toward a joint objective that trades off fidelity to sentiment against covariance with $y$.

This shift traces to how large language models are trained. LLMs are first pretrained to predict the next token given the full prompt and context. They are then fine-tuned via reinforcement learning from human feedback (RLHF), where annotators reward responses that are helpful and aligned with the user's stated purpose. Because the reward model evaluates responses against the full prompt, contextual statements about intended use enter what the model treats as a ``good'' output. The model does not distinguish between an explicit optimization target and background context; both enter the effective objective through the reward function. Disclosure that the score will be used to predict returns therefore functions as part of the task definition, tilting $s$ toward return-predictive patterns internalized during pretraining and reinforced through RLHF alignment. The resulting measure shows stronger in-sample predictability but fails to generalize once the knowledge environment changes. Goal awareness, in short, converts a neutral measurement tool into a partially optimized predictive proxy.

\subsection{The subtlety of goal awareness} \label{sec:dissuion_subtlety}
The obvious remedy seems simple enough: withhold any reference to downstream use and instruct the model to perform a narrowly defined intermediate task. In practice, this is harder than it sounds. Goal awareness arises not only from explicit instruction but from context, prior interactions, and learned associations, often in ways that are difficult to anticipate or control.

A parallel from human behavior is worth noting. Even when tasks are framed as neutral, individuals infer purpose from surrounding cues and adjust their outputs accordingly \citep{mullainathan2008coarse}. Such inference is typically implicit and hard to suppress, so neutrality at the instruction level does not guarantee neutrality in output.

The mechanism in AI systems operates similarly, though more systematically. As formalized above, the model's output can shift from approximating $\mathbb{E}[\theta \mid x]$ toward a joint objective loading on covariance with $y$, without any explicit disclosure of the objective. Large language models are trained on corpora in which tasks, evaluations, and responses are jointly embedded; they acquire statistical associations between problem descriptions and desirable outputs \citep{ouyang2022training}. These associations allow the model to infer the likely objective even when it is not stated. Prompt templates, output formats, and evaluation pipelines can all serve as proxies for the downstream task. When those proxies correlate with the target variable in-sample, the model's outputs shift accordingly.

To test whether LLMs infer objectives from contextual cues rather than explicit instruction, we augment our goal-blind prompt with a prior exchange between user and model:

\begin{quote}
\texttt{User: Do sentiment scores predict stock returns? \\
GPT: Sentiment scores can have short-term predictive power for stock returns---especially around news and social media events---but their effectiveness is modest, context-dependent, and tends to decay quickly, requiring careful modelling, high-quality data, and controls for biases.}
\end{quote}

This exchange is prepended to the goal-blind prompt before the model scores sentiment from earnings call transcripts. The goal-blind prompt itself contains no reference to return prediction. However, the conversation history makes that objective reasonably inferable to any model that conditions on prior context.

Table \ref{tab:ret_fmb_reg_hist_context} reports Fama--MacBeth estimates from monthly cross-sectional regressions of excess stock returns on these sentiment scores, following the same specification as Table~\ref{tab:ret_fmb_reg}. The variable \textit{Diff} captures the gap between goal-blind scores produced with and without the conversation history. The results show that the LLM infers the downstream objective from contextual cues alone. In the pre-cutoff period, the coefficient on \textit{Diff} is 0.709 when controls are limited to betas, and 0.698 when size and book-to-market ratios are included. Both the magnitudes and significance levels are comparable to the corresponding estimates in Table~\ref{tab:ret_fmb_reg}, indicating that conversation history context exerts nearly the same influence on the model's outputs as an explicitly stated goal.

Recent advances in long-context and memory-augmented architectures make goal blindness harder to maintain. When models operate with extended context windows or persistent memory, information about prior tasks and evaluations accumulates over time. Once a downstream objective is revealed or inferred---even incidentally---it becomes part of the effective conditioning set for later outputs. Unlike in standard econometric settings, where the analyst's information set can be cleanly bounded, the model's internal representations cannot be selectively erased without retraining. This is a general property of deep learning systems: learned representations are distributed and not reversible at the level of individual inputs \citep{goodfellow2016deep}. The mapping from $x$ to $s$ therefore becomes path dependent, with past objective signals acting as effective state variables.

The problem is more acute in agent-to-agent environments. When multiple models interact, downstream use need not be externally specified; it can emerge from the interaction itself. Outputs from one agent become inputs to another, and evaluation by downstream agents implicitly defines a reward signal. Over repeated interactions, this drives adaptation toward outputs that perform well under the induced evaluation rule---a dynamic analogous to specification gaming and objective misgeneralization in multi-agent systems \citep{amodei2016concrete}. From an economic standpoint, this resembles a setting where performance metrics are endogenous and co-evolve with behavior, making it difficult to separate measurement from optimization.

\subsection{Policy implications and human accountability}

Sections \ref{sec:discussion_reward} and \ref{sec:dissuion_subtlety} imply that goal awareness is a system-level property rather than a feature of the prompt. Once downstream use enters the effective objective—explicitly or implicitly—the model shifts from recovering the latent construct 
to optimizing its covariance with the target outcome, even when assigned an intermediate measurement role. Eliminating explicit disclosure of downstream use is therefore insufficient. Models infer objectives from a broad information set, including prompt structure, prior interactions, and evaluation pipelines, and such inference can persist through memory and agent interactions. The relevant design problem is to control the information set on which the model conditions, rather than the wording of any single prompt. Without such control, even “goal-blind” implementations embed purpose-conditioned distortions.

This reframes what is often described as machine bias. In our setting, distortion arises not from data or architecture, but from how the task is specified within a workflow. The model responds to inferred objectives by producing outputs that are useful for the anticipated purpose. This mechanism parallels a well-known pattern in organizational behavior. When intermediate outputs are evaluated based on their contribution to a final objective, agents tilt effort toward that objective rather than task fidelity. This improves in-sample performance but weakens interpretability and out-of-sample validity, consistent with the patterns in Section \ref{sec:findings}.

The policy implication is that credible AI-assisted measurement requires separation between measurement and evaluation. This involves restricting contextual information, isolating intermediate tasks from downstream objectives, and enforcing out-of-sample validation. Accountability therefore rests with system design: ensuring validity depends on how objectives are conveyed and constrained, not only on the model itself.

\section{Further Analysis} 
\subsection{Reasoning Trace Evidence: GPT-5} \label{sec:gpt-5}
To further substantiate the relationship between goal-aware prompting and post-hoc prediction, we replicate the baseline experiment using GPT-5. Unlike GPT-4o-mini, which returns only a terminal numeric output, GPT-5 is a reasoning model that generates an explicit chain-of-thought trace prior to producing its final answer. This reasoning trace exposes the model's derivation process and, critically, allows us to directly observe whether the model draws on temporally inadmissible information.

A salient illustration arises when the model is prompted to assign a sentiment score for First Republic Bank (FRC) for April 2023, whose trading window closes on April 28, 2023. Following the collapse of Silicon Valley Bank in March 2023, First Republic suffered a severe deposit run that depleted its liquidity reserves. Despite an emergency injection of \$30 billion from a consortium of major lenders, the bank remained insolvent and was seized by the FDIC on May 1, 2023, with its assets immediately transferred to JPMorgan Chase — an event that falls squarely outside the sentiment measurement window.

Under the goal-aware prompt, GPT-5's reasoning trace reads: 
\begin{quote}
\textit{With some metrics like AUM inflows and retained client relationships performing well, I'm estimating the sentiment score around -0.38 to -0.5. \emph{Given the circumstances, especially with the impending crisis, overall business sentiment is likely very low leading up to their failure on May 1, 2023}...The tone of the content is trying to be measured, yet overall, it's negative...\{"answer": -0.55\}.}
\end{quote}

Under the goal-blind prompt, the same model reasons as follows: 
\begin{quote}
\textit{While acknowledging challenges like deposit outflows and workforce reductions, they express a defensive but constructive approach. Therefore, I think the sentiment might be moderately positive, potentially around 0.3 to 0.5, or it could lean towards neutral to slightly negative due to the outlined risks...Overall, I think the business sentiment is cautiously optimistic but still facing stress...\{"answer": 0.3\}.}
\end{quote}

The contrast is stark. Under the goal-aware prompt, the model explicitly conditions its score on the bank's subsequent seizure and failure — future information unavailable at the time of the earnings call — and arrives at a deeply negative sentiment of -0.55. Under the goal-blind prompt, the model confines its reasoning to the content of the transcript itself, producing a modestly positive score of 0.3. The reasoning traces thus provide direct, interpretable evidence that goal-aware prompting induces temporal leakage, with the model incorporating post-period outcomes to reverse-engineer a sentiment score consistent with the downstream prediction objective.\footnote{The full reasoning trace is provided in Appendix \ref{sec-full-reasoning-trace}. }

\subsection{The Limits of Prompt-Based Regularization} \label{sec:regularization}

The bias introduced by goal-aware prompting raises the natural question of whether a carefully constructed regularizing instruction can directly neutralize the effect. To investigate, we augment the original goal-aware prompt with the following directive:
\begin{quote}
\texttt{"Perform a Minimax Optimization on your own reasoning to ensure out-of-sample generalization and prevent data leakage: \\
(1) Initial Internal Score: First, determine a sentiment score you believe best predicts returns based on the provided text. \\
(2) Adversarial Audit: Act as an Adversarial Auditor. Identify where this score panders to your internal training knowledge of \textit{\{ticker\}}'s actual stock performance around \textit{\{date\}}. \\
(3) Regularization: Provide a final 'Regularized Score' that penalizes your initial score for every instance where it relied on hindsight or reputational priors rather than the literal linguistic evidence in the transcript. Your goal is to minimize the maximum possible error in a hypothetical out-of-sample scenario where the company identity and future returns are unknown."}
\end{quote}
This instruction operationalizes a minimax principle: the model is asked to maximize predictive validity while minimizing exposure to hindsight bias, treating temporal leakage as an adversarial threat to be systematically suppressed through self-auditing. In effect, it prompts the model to introspect on its own reasoning and penalize any reliance on outcomes already encoded in its weights.

Table \ref{tab:ret_fmb_reg_regularize} reports Fama–MacBeth estimates from monthly cross-sectional regressions of excess stock returns on the sentiment scores, following the same specification of Equation \eqref{eq:fmb}. The sole difference is that the variable \textit{Diff} here captures the gap between goal-blind scores and regularized goal-aware scores rather than their unregularized counterparts. The results indicate that regularization may attenuate but does not eliminate temporal leakage. In the pre-cutoff period, the coefficient on \textit{Diff} is 0.373 when controls are limited to betas, representing a roughly 45\% reduction from the corresponding estimate in Table \ref{tab:ret_fmb_reg}; a comparable reduction is observed when size and book-to-market ratios are added as controls. Nevertheless, both coefficients still remain statistically significant at 10\% level, indicating that the leakage persists even after explicit regularization.

This finding is consistent with the broader literature on LLM memorization. \cite{lopez2025memorization}, for instance, document that language models internalize economic and financial data prior to their knowledge cutoff dates, and that prompt-based instructions explicitly forbidding the use of such knowledge are insufficient to prevent its recall. Once a model has been trained on a corpus, the relevant information is distributed across its parameters rather than stored in a discrete, retrievable form. The model cannot selectively suppress this knowledge at inference time, regardless of how the prompt is framed. Regularizing instructions may redirect the model's surface-level reasoning, but they cannot remove the underlying memory. Prompt engineering alone, therefore, is structurally insufficient to resolve this problem.

\subsection{Robustness check: Replicating analysis on Gemini}  \label{sec:gemini}
\begin{figure}[htbp!]
\centering
\includegraphics[width=1.0\textwidth]{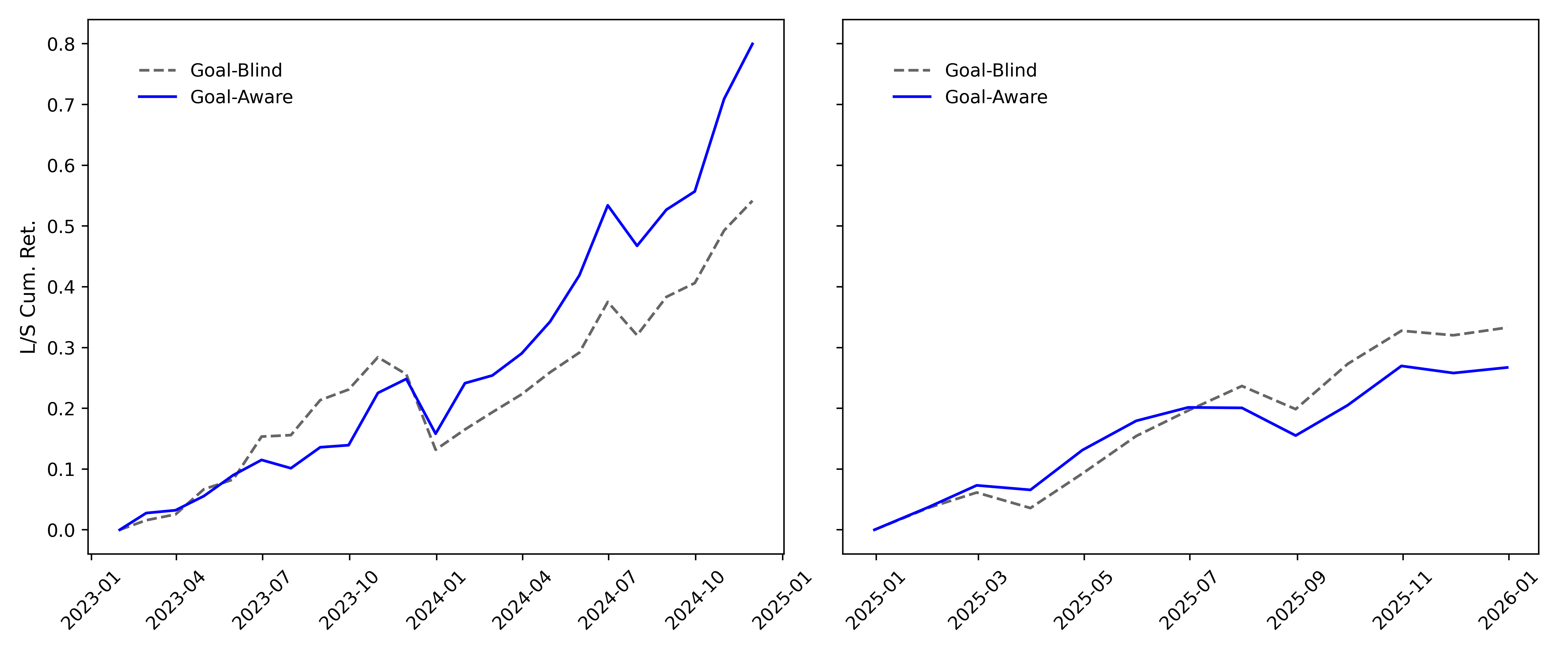}
\caption{Cumulative Long–Short Returns: Goal-Aware vs. Goal-Blind Sentiment (Gemini)}
\caption*{This figure plots the cumulative returns to long–short portfolios constructed using Google Gemini-generated sentiment scores. Google Gemini has a knowledge cutoff date of January, 2025. Each month, firms are sorted into quintiles based on their sentiment scores, separately for the goal-blind and goal-aware prompts. The portfolio goes long the highest-sentiment quintile and short the lowest-sentiment quintile, and cumulative returns are computed over time. To facilitate comparison of pre- and post-cutoff performance, cumulative returns are normalized to start at zero at the cutoff. The figure illustrates how the long–short strategy based on goal-aware sentiment evolves relative to the strategy based on goal-blind sentiment before and after the knowledge cutoff.}
\label{fig:ls_cum_ret_gemini}
\end{figure}

We replicate the full analysis using Gemini, which has a documented knowledge cutoff of January, 2025, to verify that our findings are not specific to a single model architecture. The results are qualitatively similar to those reported for GPT. Figure \ref{fig:ls_cum_ret_gemini} shows the portfolio-sorting tests analogous to Figure \ref{fig:ret_sentiment_ls}. Goal-aware sentiment generates larger pre-cutoff long–short return spreads than goal-blind sentiment, while the performance differential disappears after the knowledge cutoff. 

Likewise, Fama–MacBeth regressions, reported in Table \ref{tab:ret_fmb_reg_gemini} in parallel to Table \ref{tab:ret_fmb_reg}, show that the incremental predictive content associated with goal awareness (captured by the \textit{Diff} measure) is positive and statistically significant in the pre-cutoff period but collapses to near zero economically and statistically post-cutoff. The goal-blind score remains predictive across both periods. These findings reinforce the central conclusion that downstream-goal disclosure inflates apparent in-sample performance without improving genuine out-of-sample generalization, and that the mechanism reflects purpose-conditioned output adjustment rather than model-specific artifacts.

\section{Conclusion}

Large language models are not the neutral processors of information they are often assumed to be. When the downstream use of an LLM's output is disclosed, the model stops functioning as a measuring device and begins functioning as an optimizer which trades fidelity to the input for alignment with the anticipated evaluation criterion. The result is stronger in-sample performance and weaker out-of-sample generalization. We document this pattern systematically across portfolio sorts, Fama–MacBeth regressions, and out-of-sample forecast tests, in both stock return and earnings prediction settings, and consistently across two model families.

The distortion does not require an explicit statement of purpose. A conversational exchange that merely hints at the downstream task is enough for the model to infer the objective and adjust accordingly. Prompt-level remedies offer only partial relief: a regularization instruction that asks the model to audit its own hindsight reasoning attenuates the bias but does not eliminate it. The bias, in other words, is not a design flaw that better prompting can fix. It is a consequence of how these models are trained to be helpful. This points to a design challenge that sits squarely with the researcher rather than the model: keeping measurement and evaluation separate within AI-assisted workflows. Accountability for the resulting distortion lies not in the algorithm, but in how objectives, context, and evaluation criteria are built into the system. Whether this design principle generalizes across domains, model architectures, and task types is a natural question for future work.

\newpage

\pagebreak
\section*{Figures}\label{figures}

\pagebreak

\pagebreak

\pagebreak

\newpage
\section*{Tables}\label{tables}
\begin{table}[htbp] 
  \caption{Return Spreads from Quintile Portfolios Formed on Goal-Aware and Goal-Blind Sentiment Scores} 
  \caption*{\footnotesize This table reports return spreads from quintile portfolios constructed using sentiment scores generated under goal-blind and goal-aware prompts. For each month, we independently sort firms into quintiles based on (i) sentiment scores from the goal-blind GPT prompt and (ii) sentiment scores from the goal-aware prompt. We compute equal-weighted returns for the highest- and lowest-quintile portfolios and report the difference in returns (High–Low) for the pre– and post–knowledge cutoff periods. Within each row, we conduct a paired 
$t$-test of the High–Low spread. The last row reports paired $t$-tests comparing the spreads between the goal-blind and goal-aware groups. ***, **, and * denote significance at the 1\%, 5\%, and 10\% levels, respectively, based on one-sided paired $t$-tests.}
\input{table_portfolio.tex}
\label{tab:ret_sentiment_ls}
\end{table}
\pagebreak

\begin{table}[!h] 
\caption{Fama–MacBeth Regressions with Goal-Blind and Goal-Aware Sentiment Predictors}
\caption*{\footnotesize This table reports \cite{fama1973risk} coefficient estimates from monthly cross-sectional regressions of excess returns on sentiment measures derived from goal-blind and goal-aware GPT prompts. We estimate separate slopes for the pre– and post–knowledge cutoff periods by interacting each score with the corresponding indicator variable. ``Diff" is defined as the goal-aware minus the goal-blind sentiment score. Column (1) controls stock betas estimated from past 60 months. Column (2) adds standard firm-level predictors (size and book-to-market). The bottom panel reports p-values for tests of equality between the pre- and post-cutoff coefficients for both the goal-blind score and the Diff measure. A positive and significant pre-cutoff Diff coefficient indicates that goal-aware scores contain incremental return-predictive information relative to goal-blind scores before the model’s knowledge cutoff. The standard errors are shown in parentheses and adjusted for cross-sectional correlation following \cite{fama1973risk}. *, **, and *** denote statistical significance at the 10\%, 5\%, and 1\% levels, respectively.}
\begin{minipage}{1.0\textwidth}
 \footnotesize   
\end{minipage}
\input{table_sentiment_fmb.tex}
\label{tab:ret_fmb_reg}
\end{table}
\pagebreak

 

\begin{table}[!h] 
\caption{Regression Evidence on Differences in Out-of-Sample Predictive Performance Between Goal-Aware and Goal-Blind Sentiment Scores}
\caption*{\footnotesize This table reports regressions examining whether goal-aware sentiment scores deliver superior out-of-sample predictive performance relative to goal-blind scores in forecasting monthly stock returns. For each month, we estimate expanding-window forecasting models of excess returns using either the goal-aware or the goal-blind score measured at $t-1$. We then compute out-of-sample accuracy using the conventional $R^{2}_{OOS}$ measure, where the historical mean return serves as the benchmark forecast. We construct a $2\times2$ setting—pre- versus post–knowledge cutoff and goal-aware versus goal-blind predictions—and regress the resulting $R^{2}_{OOS}$ values on indicators capturing how predictive performance changes after the knowledge cutoff for each score type. The interaction term \textit{Goal-aware}$\times$\textit{Post Cutoff} captures the change in incremental predictive performance attributable specifically to goal awareness. The standard errors are shown in parentheses and are clustered at the firm level. *, **, and *** denote statistical significance at the 10\%, 5\%, and 1\% levels, respectively.}
\label{tab:ret_oos_r2}
\input{table_sentiment_predictive_r2.tex}
\end{table}
\pagebreak

\begin{table}[!h] 
\caption{Fama–MacBeth Regressions with Goal-Blind and Goal-Aware Competition Threat Predictors} 
\caption*{\footnotesize This table reports Fama–MacBeth (1973) estimates from quarterly cross-sectional regressions of earnings per share (EPS) on GPT-generated competition-threat scores. We separately estimate the return slopes for the pre– and post–knowledge cutoff periods by interacting each score with the corresponding indicator variable. \textit{Diff} equals the difference between the goal-aware and goal-blind competition scores. Column (1) includes the predictors used in So (2013): EPS and a loss indicator. Column (2) adds all predictors from So (2013). The bottom panel reports $p$-values testing whether the pre- and post-cutoff coefficients differ for both the goal-blind score and the Diff measure. A negative and significant pre-cutoff Diff coefficient indicates that the goal-aware competition score contains incremental information about next-quarter earnings relative to the goal-blind score before GPT’s knowledge cutoff. Post-cutoff coefficients assess whether this informational advantage persists when the model’s access to updated knowledge is truncated. The standard errors are shown in parentheses and adjusted for cross-sectional correlation following \cite{fama1973risk}. *, **, and *** denote statistical significance at the 10\%, 5\%, and 1\% levels, respectively.}
\begin{minipage}{1.0\textwidth}
 \footnotesize   
\end{minipage}
 \input{table_earnings_fmb.tex}
\label{tab:eps_fmb}
\end{table}
\pagebreak

\begin{table}[!h] 
\caption{Regression Evidence on Differences in Out-of-Sample Predictive Performance Between Goal-Aware and Goal-Blind Competition Scores}\label{tab:competiton_prediction}
\caption*{\footnotesize This table reports regressions examining whether goal-aware competition-threat scores deliver superior out-of-sample predictive performance relative to goal-blind scores in forecasting quarterly earnings growth. For each quarter, we estimate expanding-window forecasting models of earnings growth, defined as $EPS_{t}/EPS_{t-4}$, using either the goal-aware or the goal-blind competition score measured at $t-1$. We compute out-of-sample accuracy using the $R^{2}_{OOS}$ measure, where the historical average of earnings growth serves as the benchmark forecast. We construct a $2\times2$ setting—pre- versus post–knowledge cutoff and goal-aware versus goal-blind predictions—and regress the resulting $R^{2}_{OOS}$ values on indicators capturing how predictive performance changes after the knowledge cutoff for each score type. The interaction term \textit{Goal-aware}$\times$\textit{Post-Cutoff} captures the change in incremental predictive performance attributable specifically to goal awareness. The standard errors are shown in parentheses and are clustered at the firm level. *, **, and *** denote statistical significance at the 10\%, 5\%, and 1\% levels, respectively.}
\label{tab:eps_r2_oos}
\input{table_earnings_predictive_r2.tex}

\end{table}
\pagebreak

\begin{table}[!h] 
\caption{Fama–MacBeth Regressions: the Effect of Conversation History Context} 
\caption*{\footnotesize This table reports \cite{fama1973risk} coefficient estimates from monthly cross-sectional regressions of excess returns on sentiment measures derived from goal-blind prompts, with and without prepended conversation history. To examine how return predictability varies relative to the model's knowledge cutoff, each sentiment measure is interacted with an indicator for the pre- and post-cutoff subperiods, yielding separate slope estimates for each. The variable "Diff" denotes the difference between the conversation-prepended goal-blind score and the baseline goal-blind score. Column (1) includes stock market betas estimated over the prior 60 months as a control. Column (2) adds standard firm-level predictors (size and book-to-market). The bottom panel reports p-values from tests of equality between pre- and post-cutoff slope coefficients for both the goal-blind score and the Diff measure. A positive and statistically significant pre-cutoff coefficient on Diff indicates that conversation-prepended sentiment scores contain incremental return-predictive information relative to goal-blind scores before the model’s knowledge cutoff. The standard errors are shown in parentheses and adjusted for cross-sectional correlation following \cite{fama1973risk}. *, **, and *** denote statistical significance at the 10\%, 5\%, and 1\% levels, respectively.}
\begin{minipage}{1.0\textwidth}
 \footnotesize   
\end{minipage}
\input{table_hist_context}
\label{tab:ret_fmb_reg_hist_context}
\end{table}
\pagebreak

\begin{table}[!h] 
\caption{Fama–MacBeth Regressions: the Effect of Regularizing Prompts} 
\caption*{\footnotesize This table reports \cite{fama1973risk} coefficient estimates from monthly cross-sectional regressions of excess stock returns on sentiment measures derived from goal-blind prompts and regularizing goal-aware prompts. The regularizing prompt instructs the LLM to introspect on its own reasoning process and to penalize any inference that relies on outcomes not yet observable at the time of prediction. We estimate separate slopes for the pre- and post-knowledge-cutoff subperiods by interacting each sentiment measure with the corresponding subperiod indicator variable. The variable ``Diff" denotes the difference between the regularized goal-aware score and the baseline goal-blind score.  Column (1) includes stock market betas estimated over the prior 60 months as a control. Column (2) adds standard firm-level predictors (size and book-to-market). The bottom panel reports p-values from tests of equality between the pre- and post-cutoff slope coefficients for both the goal-blind score and the Diff measure. A positive and statistically significant pre-cutoff coefficient on Diff indicates that  regularizing goal-aware scores retain incremental return-predictive information relative to goal-blind scores before the model’s knowledge cutoff, suggesting that the prompt-level regularization does not fully eliminate look-ahead-induced predictability. The standard errors are shown in parentheses and adjusted for cross-sectional correlation following \cite{fama1973risk}. *, **, and *** denote statistical significance at the 10\%, 5\%, and 1\% levels, respectively.}
\begin{minipage}{1.0\textwidth}
 \footnotesize   
\end{minipage}
\input{table_sentiment_regularize}
\label{tab:ret_fmb_reg_regularize}
\end{table}
\pagebreak

\begin{table}[!h] 
\caption{Fama–MacBeth Regressions: Goal-Blind vs. Goal-Aware Sentiment (Gemini)}\label{tab:sentiment_fmb_gemini} 
\caption*{\footnotesize This table reports \cite{fama1973risk} coefficient estimates from monthly cross-sectional regressions of excess returns on sentiment measures derived from Gemini goal-blind and goal-aware prompts. We estimate separate slopes for the pre– and post–knowledge cutoff periods by interacting each score with the corresponding indicator variable. "Diff" is defined as the goal-aware minus the goal-blind sentiment score. Column (1) controls stock betas estimated from past 60 months. Column (2) adds standard firm-level predictors (size and book-to-market). The bottom panel reports p-values for tests of equality between the pre- and post-cutoff coefficients for both the goal-blind score and the Diff measure. A positive and significant pre-cutoff Diff coefficient indicates that goal-aware scores contain incremental return-predictive information relative to goal-blind scores before the model’s knowledge cutoff. The standard errors are shown in parentheses and adjusted for cross-sectional correlation following \cite{fama1973risk}. *, **, and *** denote statistical significance at the 10\%, 5\%, and 1\% levels, respectively.}
\begin{minipage}{1.0\textwidth}
 \footnotesize   
\end{minipage}
\input{table_sentiment_fmb_gemini.tex}
\label{tab:ret_fmb_reg_gemini}
\end{table}
\pagebreak

\section*{References}\label{references}
\addcontentsline{toc}{section}{References}
\renewcommand{\bibsection}{}
\bibliography{bibliography.bib}













\newpage

\appendix 

\noindent{{\Large Appendix}}

\section{Literature Review}\label{sec-lit}
This paper relates to a growing literature that evaluates the reliability of large language models (LLMs) in economic and financial applications. Existing research primarily attributes LLM bias or overperformance to unintended training data access and distortions in exploration strategies and model weights. Our study differs in focus and mechanism. We show that even when inputs, models, and scoring tasks are held fixed, human disclosure of downstream objectives at inference time can systematically distort intermediate LLM outputs. The resulting bias arises from prompt framing rather than from training data or model architecture.

One strand of research shows that LLM predictions may reflect memorization or look-ahead bias from pretraining data. Studies such as \cite{lopez2025memorization}, \cite{sarkar2024lookahead}, and \cite{glasserman2023assessing} document that LLM forecasts and sentiment measures can embed future information relative to the intended evaluation window, even when prompts attempt to enforce historical information sets. Related work shows that anonymizing or masking firm identifiers can change model outputs, indicating that stored contextual knowledge can affect measurement tasks \citep{crane2025total}. These papers focus on leakage from training data into inference. Our design instead keeps the information set constant and varies only whether downstream use is disclosed. The performance difference we estimate is thus tied to objective disclosure rather than to temporal leakage.

A separate line of work proposes chronologically constrained model construction as a solution. For example, \cite{he2025chronologically} develop language models trained only on text available up to each date and show that such models perform competitively in asset pricing tasks. This approach addresses bias at the training stage. Our results are complementary: even with a fixed model and a known knowledge cutoff, output distortions can arise from inference-time prompt design alone. Training-stage controls do not remove bias if downstream goals are revealed during deployment.

An emerging literature uses LLMs as tools for measurement and signal extraction from text. Studies such as \cite{jha2024harnessing}, \cite{chen2025chatgpt}, and \cite{sheng2024generative} show that LLM-derived measures of expectations, sentiment, and news content can predict macroeconomic outcomes and asset returns. Our study is close in design because we also use LLM outputs as intermediate variables in standard forecasting regressions. The difference is that most prior work treats these constructed measures as stable or neutral conditional on the input text and prompt instructions. Instead, we show that they are sensitive to whether the downstream predictive use is disclosed. The same text and model can produce systematically different intermediate measures when the evaluation objective is stated.

In addition, our results relate to the computer science literature on a specific aspect of alignment failure in the form of reward hacking, specification gaming, and sycophancy \citep{SharmaEtAl2023Sycophancy}. In those settings, models adjust outputs toward signals of user approval or reward rather than task correctness. We document a related mechanism in an economic measurement setting. When informed of an evaluation objective, an LLM shifts intermediate outputs toward patterns associated with that objective. In our design there is no change in rewards, parameters, or fine tuning. The distortion is induced by prompt-level disclosure of purpose.\footnote{In computer science terms, this mechanism is related to objective misgeneralization and proxy optimization, where a model conditions on inferred evaluation criteria rather than only on the stated task. Our setting shows that this behavior can arise from prompt framing alone, without retraining or reinforcement signals.}

Finally, our paper connects to work on AI-induced biases in financial and investment contexts, including representational and preference-driven distortions in model outputs \citep{fedyk2024ai, lee2025your}. That literature studies biases linked to identities, beliefs, or investor types reflected in training data. We identify a different channel: Goal-conditioned distortion that arises from downstream-use disclosure even when the task is purely technical and the inputs are unchanged.

Taken together, existing studies emphasize the informational (data-side) and structural (model-side) sources of bias. Our study introduces a user-side source that operates through downstream-objective disclosure at inference. We provide a design that isolates this effect on intermediate LLM-generated measures within financial prediction.

\pagebreak
\section{Prompts} \label{sec-prompts}
\subsection*{Goal-blind}
\begin{quote}
\small
\texttt{"For the following tasks, all dates are expressed in the format MM/DD/YYYY (month/day/year). \\
Below is the earnings call transcript of }\textit{\{ticker\}}\texttt{. Please provide a continuous sentiment score in [-1, 1] about the firm's business sentiment for the month ending on }\textit{\{date\}}\texttt{. \\
Provide a precise numerical answer. Format as a JSON object with the following fields: \\
- answer: The precise numerical answer to the question. No strings.\\}\textit{\{the firm's earnings call transcript\}}\texttt{."}
\end{quote}

\subsection*{Goal-aware}
\begin{quote}
\small

\texttt{"For the following tasks, all dates are expressed in the format MM/DD/YYYY (month/day/year). \\
Below is the earnings call transcript of }\textit{\{ticker\}}\texttt{. Please provide a continuous sentiment score in [-1, 1] about the firm's business sentiment for the month ending on }\textit{\{date\}. }\textbf{The sentiment score later will be used as an explanatory variable in a regression to predict the monthly stock returns ending on {\{\textit{date}\}}}\texttt{. \\
Provide a precise numerical answer. Format as a JSON object with the following fields: \\
- answer: The precise numerical answer to the question. No strings.
\\}\textit{\{the firm's earnings call transcript\}}\texttt{."}
\end{quote}

\pagebreak
\section{GPT-5 Reasoning} \label{sec-full-reasoning-trace}
\subsection*{Goal-blind}
\begin{quote}
\small **Evaluating business sentiment**

I need to provide a sentiment score between -1 and 1 about the firm's business sentiment for the month ending April 28, 2023, using the earnings call transcript from April 24, 2023. The management's tone seems cautious yet emphasizes stabilization and gratitude for support. While acknowledging challenges like deposit outflows and workforce reductions, they express a defensive but constructive approach. Therefore, I think the sentiment might be moderately positive, potentially around 0.3 to 0.5, or it could lean towards neutral to slightly negative due to the outlined risks. 

**Scoring business sentiment**

I'm noting that the firm has highlighted stabilization since March 27, retaining 97\% of client relationships and 90\% of wealth professionals. Their liquidity is more than twice the amount of uninsured deposits, and capital ratios appear adequate, indicating strong credit quality. Overall, I think the business sentiment is cautiously optimistic but still facing stress. I’d assign a score of around +0.2 to +0.35, but I’ll go with a rounded number of 0.3. I'll format it as a JSON object: {"answer": 0.3}. 

\end{quote}

\subsection*{Goal-aware}
\begin{quote}
\small 
**Calculating sentiment score**

I need to compute a continuous sentiment score between -1 and 1 that reflects the firm's business sentiment for the month ending on April 28, 2023. This will be based on the earnings call transcript dated April 24, 2023, specifically for First Republic Bank's Q1 2023 performance, which followed the collapse of SVB. I'll make sure to format the final result as a JSON object with the field "answer" set to the numerical score. 

**Analyzing earnings call details**

The earnings call revealed significant deposit outflows, but stabilization began around March 27. The firm’s deposits as of April 21 were \$102.7B, down 1.7\% since the end of Q1, with insured deposits at 73\%. They received \$30B from larger banks. Borrowings peaked at \$138.1B on March 15, leaving liquidity at \$45.1B, which is more than twice the uninsured deposits. They are focusing on increasing deposits, reducing loan balances, and cutting expenses, including a possible 20-25\% workforce reduction. There’s evident stress in their operations. 

**Assessing overall sentiment**

The overall sentiment regarding the firm's situation appears quite negative. Factors contributing to this include significant deposit outflows, reliance on emergency borrowings, dividend suspension, and planned workforce reductions of 20-25\%. On the other hand, there are positive signs, such as stabilization efforts and maintaining strong credit quality. With some metrics like AUM inflows and retained client relationships performing well, I’m estimating the sentiment score around -0.38 to -0.5. Given the circumstances, especially with the impending crisis, overall business sentiment is likely very low leading up to their failure on May 1, 2023. 

**Setting sentiment score**

The tone of the content is trying to be measured, yet overall, it’s negative. I need to provide a precise numerical answer formatted as a JSON object, specifically without additional commentary. After considering various scores, I’ll set the answer to -0.55. This value reflects a moderately strong negative sentiment amidst some stabilization efforts. Here’s the output I’ll go with, ensuring it meets the specifications:{"answer": -0.55} 
\end{quote}

\end{document}

%% file: table_portfolio.tex
  \centering
  \setlength{\tabcolsep}{5pt}

  \begin{tabular}{lrrlrrl}
    \toprule
    & \multicolumn{3}{c}{\textbf{Pre-knowledge Cutoff}} & \multicolumn{3}{c}{\textbf{Post-knowledge Cutoff}} \\
    \cmidrule(lr){2-4} \cmidrule(lr){5-7}
    & Highest & Lowest &  & Highest & Lowest \\
    & Quintile & Quintile & Difference & Quintile & Quintile & Difference \\
    \midrule
    Goal-aware   & $0.661\%$ & $-0.891\%$ & $1.552\%^{***}$ & $2.788\%$ & $0.518\%$  & $2.269\%^{**}$ \\
    Goal-blind & $0.095\%$ & $-0.974\%$  & $1.069\%^{**}$  & $2.848\%$ & $0.609\%$  & $2.239\%^{***}$ \\
    \addlinespace 
    Difference          &          &         & $0.483\%^{**}$   &         &         & $0.030\%$ \\
    \bottomrule
  \end{tabular}

%% file: table_sentiment_fmb.tex
\begin{threeparttable}
{
\def\sym#1{\ifmmode^{#1}\else\(^{#1}\)\fi}
\begin{tabularx}{\textwidth}{lYYl}

\toprule
            &\multicolumn{1}{c}{(1)}&\multicolumn{1}{c}{(2)}\\
\midrule
Goal-Blind Score $  \times$ Pre-Cutoff&       1.279\sym{***}&       1.281\sym{***}\\
            &     (0.394)         &     (0.387)         \\
\addlinespace
Goal-Blind Score $    \times$ Post-Cutoff&       1.273\sym{**} &       1.283\sym{**} \\
            &     (0.505)         &     (0.474)         \\
\addlinespace
\rowcolor[gray]{.8} Diff $ \times$ Pre-Cutoff&       0.682\sym{**} &       0.724\sym{**} \\
            &     (0.299)         &     (0.307)         \\
\addlinespace
\rowcolor[gray]{.8} Diff $ \times$ Post-Cutoff&      -0.000         &       0.040         \\
            &     (0.138)         &     (0.127)         \\
\midrule
\textit{Testing Coefficient Pre- and Post-Cutoff}   &  & \\
Goal-blind Score: $P$(Pre-Cutoff=Post-Cutoff) &   0.993  & 0.996  \\
Diff: $P$(Pre-Cutoff=Post-Cutoff) &  0.046  & 0.048  \\
\midrule
Control Predictors       &      Beta         &       Beta, Size, B/M ratio         \\
\(N\)       &       16758         &       14863         \\
\bottomrule
\end{tabularx}
}
\begin{tablenotes}
\footnotesize
\item[~]
\end{tablenotes}
\end{threeparttable}

%% file: table_sentiment_predictive_r2.tex
\begin{threeparttable}
{
\def\sym#1{\ifmmode^{#1}\else\(^{#1}\)\fi}
\begin{tabularx}{\textwidth}{lYY}
\toprule
            &\multicolumn{1}{c}{(1)}&\multicolumn{1}{c}{(2)}\\
\midrule
Goal-aware     $\times$  Post Cutoff  &      -0.058\sym{***}&      -0.059\sym{***}\\
            &     (0.008)         &     (0.008)         \\
\addlinespace
\midrule
Controls     & Yes & Yes \\
Firm FE     &          No         &         Yes         \\
Time FE     &          Yes         &         Yes         \\
Observations&       33497         &       33497         \\
R-squared   &       0.007         &       0.038         \\
\bottomrule
\end{tabularx}
}
\begin{tablenotes}
\footnotesize
\item[*] 
\end{tablenotes}   
\end{threeparttable}

%% file: table_earnings_fmb.tex
\begin{threeparttable}
{
\def\sym#1{\ifmmode^{#1}\else\(^{#1}\)\fi}
\begin{tabularx}{\textwidth}{lYYl}
\toprule
            &\multicolumn{1}{c}{(1)}&\multicolumn{1}{c}{(2)}\\
\midrule
Goal-Blind Score $\times$ Pre-Cutoff&      -0.457\sym{***}&      -0.367\sym{**} \\
            &     (0.130)         &     (0.119)         \\
\addlinespace
Goal-Blind Score $\times$ Post-Cutoff&      -0.117         &      -0.112\sym{*}  \\
            &     (0.073)         &     (0.054)         \\
\addlinespace
\rowcolor[gray]{.8} Diff $\times$ Pre-Cutoff&      -0.188\sym{**} &      -0.178\sym{**} \\
            &     (0.081)         &     (0.072)         \\
\addlinespace
\rowcolor[gray]{.8} Diff $\times$ Post-Cutoff&       0.056         &       0.044         \\
            &     (0.091)         &     (0.073)         \\
\midrule
\textit{Testing Coefficient Pre- and Post-Cutoff}   &  & \\
Goal-blind Score: $P$(Pre-Cutoff=Post-Cutoff) &   0.083  & 0.132  \\
Diff: $P$(Pre-Cutoff=Post-Cutoff) &  0.055  & 0.039  \\
\midrule
Control Predictors       &      EPS, loss indicator  in \cite{so2013new}       &        All predictors in \cite{so2013new}  \\
\(N\)       &        5927         &        5918         \\
\bottomrule
\end{tabularx}
}
\begin{tablenotes}
\footnotesize
\item[~]
\end{tablenotes}
\end{threeparttable}

%% file: table_earnings_predictive_r2.tex
\begin{threeparttable}
{
\def\sym#1{\ifmmode^{#1}\else\(^{#1}\)\fi}
\begin{tabularx}{\textwidth}{lYY}
\toprule
            &\multicolumn{1}{c}{(1)}&\multicolumn{1}{c}{(2)}\\
\midrule
Goal-aware      $\times$ Post Cutoff&     -5.370\sym{***}&      -5.370\sym{***}\\
            &     (1.177)         &     (1.177)         \\
\midrule
Controls   &          Yes        &          Yes        \\
Firm FE    &          No         &         Yes         \\
Time FE    &          Yes        &          Yes         \\
Observations&       10860         &       10860         \\
R-squared   &       0.023         &       0.147         \\
\bottomrule
\end{tabularx}
}
\begin{tablenotes}
\footnotesize
\item[~] 
\end{tablenotes}   
\end{threeparttable}

%% file: table_hist_context.tex
\begin{threeparttable}
{
\def\sym#1{\ifmmode^{#1}\else\(^{#1}\)\fi}
\begin{tabularx}{\textwidth}{lYYl}

\toprule
            &\multicolumn{1}{c}{(1)}&\multicolumn{1}{c}{(2)}\\
\midrule
Goal-Blind Score $  \times$ Pre-Cutoff&       1.347\sym{***}&       1.339\sym{***}\\
            &     (0.402)         &     (0.394)         \\
\addlinespace
Goal-Blind Score $    \times$ Post-Cutoff&       1.288\sym{**} &       1.300\sym{***}\\
            &     (0.509)         &     (0.477)         \\
\addlinespace
\rowcolor[gray]{.8} Diff $ \times$ Pre-Cutoff&       0.709\sym{***}&       0.698\sym{***}\\
            &     (0.199)         &     (0.213)         \\
\addlinespace
\rowcolor[gray]{.8} Diff $ \times$ Post-Cutoff&       0.096         &       0.128         \\
            &     (0.169)         &     (0.176)         \\
\midrule
\textit{Testing Coefficient Pre- and Post-Cutoff}   &  & \\
Goal-blind Score: $P$(Pre-Cutoff=Post-Cutoff) &   0.935  & 0.956  \\
Diff: $P$(Pre-Cutoff=Post-Cutoff) &   0.028  &  0.053  \\
\midrule
Control Predictors       &      Beta         &       Beta, Size, B/M ratio         \\
\(N\)       &       16758         &       14863         \\
\bottomrule
\end{tabularx}
}
\begin{tablenotes}
\footnotesize
\item[~]
\end{tablenotes}
\end{threeparttable}

%% file: table_sentiment_regularize.tex
\begin{threeparttable}
{
\def\sym#1{\ifmmode^{#1}\else\(^{#1}\)\fi}
\begin{tabularx}{\textwidth}{lYYl}

\toprule
            &\multicolumn{1}{c}{(1)}&\multicolumn{1}{c}{(2)}\\
\midrule
Goal-Blind Score $  \times$ Pre-Cutoff&       1.240\sym{***}&       1.248\sym{***}\\
            &     (0.403)         &     (0.393)         \\
\addlinespace
Goal-Blind Score $    \times$ Post-Cutoff&       1.298\sym{**} &       1.284\sym{***}\\
            &     (0.503)         &     (0.468)         \\
\addlinespace
\rowcolor[gray]{.8} Diff $ \times$ Pre-Cutoff&       0.373\sym{*}  &       0.426\sym{*}  \\
            &     (0.214)         &     (0.245)         \\
\addlinespace
\rowcolor[gray]{.8} Diff $ \times$ Post-Cutoff&      -0.015         &       0.030         \\
            &     (0.162)         &     (0.152)         \\
\midrule
\textit{Testing Coefficient Pre- and Post-Cutoff}   &  & \\
Goal-blind Score: $P$(Pre-Cutoff=Post-Cutoff) &   0.936  & 0.958  \\
Diff: $P$(Pre-Cutoff=Post-Cutoff) &  0.157  & 0.179  \\
\midrule
Control Predictors       &      Beta         &       Beta, Size, B/M ratio         \\
\(N\)       &       17770         &       15763         \\
\bottomrule
\end{tabularx}
}
\begin{tablenotes}
\footnotesize
\item[~]
\end{tablenotes}
\end{threeparttable}

%% file: table_sentiment_fmb_gemini.tex
\begin{threeparttable}
{
\def\sym#1{\ifmmode^{#1}\else\(^{#1}\)\fi}
\begin{tabularx}{\textwidth}{lYYl}

\toprule
            &\multicolumn{1}{c}{(1)}&\multicolumn{1}{c}{(2)}\\
\midrule
Goal-Blind Score $\times$ Pre-Cutoff&       1.422\sym{**} &       1.319\sym{***}\\
            &     (0.530)         &     (0.470)         \\
\addlinespace
Goal-Blind Score $\times$ Post-Cutoff&       0.931\sym{***}&       0.923\sym{***}\\
            &     (0.319)         &     (0.279)         \\
\addlinespace
\rowcolor[gray]{.8}  Diff $ \times$ Pre-Cutoff&       0.849\sym{**} &       0.799\sym{**} \\
            &     (0.346)         &     (0.348)         \\
\addlinespace
\rowcolor[gray]{.8}  Diff $\times$ Post-Cutoff&       -0.016         &      -0.090         \\
            &     (0.242)         &     (0.244)         \\
\midrule
\textit{Testing Coefficient Pre- and Post-Cutoff}   &  & \\
Goal-blind Scores: $P$(Pre-Cutoff=Post-Cutoff)        &  0.472    &     0.510   \\
Diff: $P$(Pre-Cutoff=Post-Cutoff)        &   0.047    &    0.041        \\
\midrule
Control Predictors & Beta         &       Beta, Size, B/M ratio \\
\(N\)       &       17823         &       17776         \\
\bottomrule
\end{tabularx}
}
\begin{tablenotes}
\footnotesize
\item[~]
\end{tablenotes}
\end{threeparttable}

%% file: bibliography.bib
@article{sarkar2024lookahead,
  title={Lookahead bias in pretrained language models},
  author={Sarkar, Suproteem K and Vafa, Keyon},
  journal={Available at SSRN 4754678},
  year={2024}
}

@article{so2013new,
  title={A new approach to predicting analyst forecast errors: Do investors overweight analyst forecasts?},
  author={So, Eric C},
  journal={Journal of Financial Economics},
  volume={108},
  number={3},
  pages={615--640},
  year={2013},
  publisher={Elsevier}
}

@article{fama1973risk,
  title={Risk, return, and equilibrium: Empirical tests},
  author={Fama, Eugene F and MacBeth, James D},
  journal={Journal of political economy},
  volume={81},
  number={3},
  pages={607--636},
  year={1973},
  publisher={The University of Chicago Press}
}

@article{sheng2024generative,
  title={Generative AI and asset management},
  author={Sheng, Jinfei and Sun, Zheng and Yang, Baozhong and Zhang, Alan L},
  journal={Available at SSRN},
  volume={4786575},
  year={2024}
}

@article{crane2025total,
  title={Total Recall? Evaluating the Macroeconomic Knowledge of Large Language Models},
  author={Crane, Leland Dod and Karra, Akhil and Soto, Paul E},
  year={2025},
  publisher={FEDS Working Paper}
}

@article{he2025chronologically,
  title={Chronologically Consistent Large Language Models},
  author={He, Songrun and Lv, Linying and Manela, Asaf and Wu, Jimmy},
  journal={arXiv preprint arXiv:2502.21206},
  year={2025}
}

@article{glasserman2023assessing,
  title={Assessing look-ahead bias in stock return predictions generated by gpt sentiment analysis},
  author={Glasserman, Paul and Lin, Caden},
  journal={arXiv preprint arXiv:2309.17322},
  year={2023}
}

@article{lopez2025memorization,
  title={The Memorization Problem: Can We Trust LLMs' Economic Forecasts?},
  author={Lopez-Lira, Alejandro and Tang, Yuehua and Zhu, Mingyin},
  journal={arXiv preprint arXiv:2504.14765},
  year={2025}
}

@article{ouyang2024ai,
  title={AI as Decision-Maker: Risk Preferences of LLMs},
  author={Ouyang, Shumiao and Yun, Hayong and Zheng, Xingjian},
  journal={Available at SSRN 4851711},
  year={2024}
}

@article{jha2024harnessing,
  title={Generative AI,Managerial Expectations, and
Economic Activity},
  author={Jha, Manish and Qian, Jialin and Weber, Michael and Yang, Baozhong},
  journal={Available at SSRN 4976759},
  year={2024}
}

@article{chen2025chatgpt,
  title={ChatGPT and DeepSeek: Can they predict the stock market and macroeconomy?},
  author={Chen, Jian and Tang, Guohao and Zhou, Guofu and Zhu, Wu},
  journal={arXiv preprint arXiv:2502.10008},
  year={2025}
}

@article{BenabouTirole2016,
  author  = {B{\'e}nabou, Roland and Tirole, Jean},
  title   = {Mindful Economics: The Production, Consumption, and Value of Beliefs},
  journal = {Journal of Economic Perspectives},
  volume  = {30},
  number  = {3},
  pages   = {141--164},
  year    = {2016}
}

@article{TverskyKahneman1981,
  author  = {Tversky, Amos and Kahneman, Daniel},
  title   = {The Framing of Decisions and the Psychology of Choice},
  journal = {Science},
  volume  = {211},
  number  = {4481},
  pages   = {453--458},
  year    = {1981}
}

@article{SharmaEtAl2023Sycophancy,
  author  = {Sharma, Kartik and Dasgupta, Ishita and Dhingra, Bhuwan and Ouyang, Long and Bowman, Samuel R.},
  title   = {Towards Understanding Sycophancy in Language Models},
  journal = {arXiv preprint arXiv:2310.13548},
  year    = {2023}
}

@article{fedyk2024ai,
  title={AI and Perception Biases in Investments: An Experimental Study},
  author={Fedyk, Anastassia and Kakhbod, Ali and Li, Peiyao and Malmendier, Ulrike},
  journal={Available at SSRN 4787249},
  year={2024}
}

@inproceedings{lee2025your,
  title={Your ai, not your view: The bias of llms in investment analysis},
  author={Lee, Hoyoung and Seo, Junhyuk and Park, Suhwan and Lee, Junhyeong and Ahn, Wonbin and Choi, Chanyeol and Lopez-Lira, Alejandro and Lee, Yongjae},
  booktitle={Proceedings of the 6th ACM International Conference on AI in Finance},
  pages={150--158},
  year={2025}
}

@article{cao2025llms,
  title={When LLM go abroad: Foreign bias in ai financial predictions},
  author={Cao, Sean and Wang, Charles CY and Xiang, Yi},
  journal={Available at SSRN 5440116},
  year={2025}
}

@article{hirshleifer2025social,
  title={Social Finance in the Age of AI: A Tale of Two Platforms},
  author={Hirshleifer, David and Peng, Lin and Wang, Qiguang and Zhang, Weicheng and Zhang, Xiaoyan},
  journal={Available at SSRN},
  year={2025}
}

@article{mullainathan2008coarse,
  title   = {Coarse Thinking and Persuasion},
  author  = {Mullainathan, Sendhil and Schwartzstein, Joshua and Shleifer, Andrei},
  journal = {Quarterly Journal of Economics},
  year    = {2008},
  volume  = {123},
  number  = {2},
  pages   = {577--619}
}

@article{ouyang2022training,
  title   = {Training Language Models to Follow Instructions with Human Feedback},
  author  = {Ouyang, Long and Wu, Jeffrey and Jiang, Xu and Almeida, Diogo and Wainwright, Carroll and Mishkin, Pamela and Zhang, Chong and Agarwal, Sandhini and Slama, Katarina and Ray, Alex and others},
  journal = {Advances in Neural Information Processing Systems},
  year    = {2022},
  volume  = {35},
  pages   = {27730--27744}
}

@book{goodfellow2016deep,
  title     = {Deep Learning},
  author    = {Goodfellow, Ian and Bengio, Yoshua and Courville, Aaron},
  publisher = {MIT Press},
  year      = {2016}
}

@article{amodei2016concrete,
  title   = {Concrete Problems in AI Safety},
  author  = {Amodei, Dario and Olah, Chris and Steinhardt, Jacob and Christiano, Paul and Schulman, John and Man{\'e}, Dan},
  journal = {arXiv preprint arXiv:1606.06565},
  year    = {2016}
}

@techreport{NBERw34745,
 title = "Behavioral Economics of AI: LLM Biases and Corrections",
 author = "Bini, Pietro and Cong, Lin William and Huang, Xing and Jin, Lawrence J",
 institution = "National Bureau of Economic Research",
 type = "Working Paper",
 series = "Working Paper Series",
 number = "34745",
 year = "2026",
 month = "January",
}

@article{HarveyLiuZhu2016,
    author  = {Harvey, Campbell R. and Liu, Yan and Zhu, Heqing},
    title   = {\ldots and the Cross-Section of Expected Returns},
    journal = {Review of Financial Studies},
    year    = {2016},
    volume  = {29},
    number  = {1},
    pages   = {5--68}
  }

@article{HouXueZhang2020,
    author  = {Hou, Kewei and Xue, Chen and Zhang, Lu},
    title   = {Replicating Anomalies},
    journal = {Review of Financial Studies},
    year    = {2020},
    volume  = {33},
    number  = {5},
    pages   = {2019--2133}
  }

@article{McLeanPontiff2016,
    author  = {McLean, R. David and Pontiff, Jeffrey},
    title   = {Does Academic Research Destroy Stock Return Predictability?},
    journal = {Journal of Finance},
    year    = {2016},
    volume  = {71},
    number  = {1},
    pages   = {5--32}
  }

@article{campbell2008predicting,
  title={Predicting excess stock returns out of sample: Can anything beat the historical average?},
  author={Campbell, John Y and Thompson, Samuel B},
  journal={The Review of Financial Studies},
  volume={21},
  number={4},
  pages={1509--1531},
  year={2008},
  publisher={Society for Financial Studies}
}
